\documentclass[twoside,british,3p]{elsarticle}
\usepackage[T1]{fontenc}
\usepackage[latin9]{inputenc}
\usepackage[table,xcdraw]{xcolor}
\usepackage{pagecolor}
\usepackage{pdfcolmk}
\usepackage{units}
\usepackage{textcomp}
\usepackage{bbding}
\usepackage{amsmath}
\usepackage{amssymb}
\usepackage{graphicx}
\usepackage{setspace}
\usepackage{esvect}
\PassOptionsToPackage{normalem}{ulem}
\usepackage{ulem}
\usepackage{blindtext}
\usepackage{multicol}
\usepackage{booktabs}
\usepackage{tabularx}
\usepackage{longtable}
\usepackage{pbox}
\usepackage{csquotes}

\usepackage{caption}
\usepackage{subcaption}

\usepackage{lscape}
\usepackage{hyperref}
\usepackage{makecell}
\usepackage{multirow}

\newtheorem{theorem}{Observation}
\newcount\savedtheorem

\hypersetup{
    colorlinks=true,
    linkcolor=blue,
    filecolor=magenta,      
    urlcolor=cyan,
    pdfborderstyle={/S/U}
}

\makeatletter

\journal{Expert Systems with Applications}
\bibpunct{(}{)}{;}{a}{,}{,} 

\DeclareGraphicsExtensions{.pdf,.jpeg,.png,.eps}

\usepackage{dcolumn}   
\usepackage{bm}

\usepackage{babel}

\begin{document}

\begin{frontmatter}{}

\title{How fair were COVID-19 restriction decisions? A data-driven investigation of England using the dominance-based rough sets approach}


\author[aff1]{Edward Abel\corref{fn1}}
\ead{abel@sdu.dk}
\author[aff2,aff3]{Sajid Siraj}
\ead{s.siraj@leeds.ac.uk}

\cortext[fn1]{Room 41.58, Universitetsparken 1, University of Southern Denmark, Kolding, Denmark}

\address[aff1]{University of Southern Denmark, Denmark}
\address[aff2]{Centre for Decision Research, Leeds University Business School, Leeds, UK}
\address[aff3]{COMSATS University Islamabad, Wah Campus, Pakistan}

\begin{abstract}
During the COVID-19 pandemic, several countries have taken the approach of tiered restrictions which has remained a point of debate due to a lack of transparency. Using the dominance-based rough set approach, we identify patterns in the COVID-19 data pertaining to the UK government's tiered restrictions allocation system. These insights from the analysis are translated into "if-then" type rules, which can easily be interpreted by policy makers. The differences in the rules extracted from different geographical areas suggest inconsistencies in the allocations of tiers in these areas. We found that the differences delineated an overall north south divide in England, however, this divide was driven mostly by London. Based on our analysis, we demonstrate the usefulness of the dominance-based rough sets approach for investigating the fairness and explainabilty of decision making regarding COVID-19 restrictions. The proposed approach and analysis could provide a more transparent approach to localised public health restrictions, which can help ensure greater conformity to the public safety rules.

\end{abstract}

\begin{keyword}
    multi-criteria data analysis; dominance-based rough sets; COVID-19; data wrangling; data-driven decision making  
\end{keyword}

\end{frontmatter}

\section{Introduction\label{sec:intro}}
The pandemic caused by Coronavirus disease 2019 (COVID-19) has been affecting lives around the globe. The issue of public health and safety, looking to dampen the impacts of COVID-19, forces policy makers to restrict a lot of business and social activities involving physical interactions. However, a prolonged complete lockdown becomes infeasible due to its adverse economic and social impacts, for example, leading to supply chain problems, high unemployment rates, and increases in mental health issues \citep{Nikolopoulos2021,Bassiouni2023AdvancedRestrictions}. This makes the issue a multi-objective problem where the safety-related objectives invariably conflict with the economic objectives. Therefore, several governments tried to address this issue by introducing region-level restrictions based on the health and safety risk faced by each region. For example, the UK government announced a tiered restriction system in September 2020 \citep{DepartmentOfHealthAndSocialCare}, as opposed to its previous country-wide restrictions policy. Although this appeared to be a more sensible approach than applying a nation-wide lockdown, some local authorities found this unsatisfactory, due to a lack of sufficient transparency in assigning tiers. Although the government published the set of criteria utilised to determine tier assignments, they did not publish the framework or rules that were used for actual assignment of tiers. As a result, the allocation of tiers during the pandemic has been questioned due to a lack of transparency, leading to accusations of unfair decision making. This study investigates if and how regional inequalities were reflected in the assignment of tiered restrictions during the second wave of the COVID-19 pandemic. Our paper aims to demonstrate the use of a novel Dominance-based Rough Sets Approach (DRSA) to investigate the potential inequalities in policies regarding tiered restrictions. 

In this paper, we investigate the potentially different treatment for different geographical areas, specifically focusing on the North-South divide in England. For this purpose, we propose the use of DRSA for understanding the underlying rules that were applied in different geographical areas, and for assessing the level of inconsistencies in assigning those tiered restrictions.

After acquiring and pre-processing data from various disparate sources, we extracted distinct if-then rule sets for different geographical areas and compared these rule sets to identify any inconsistencies. Our analysis found a north south divide, however, the divide was driven mostly by London.

The structure of the rest of the paper is as follows. Section \ref{sec:background} explores literature starting on the tiered restrictions and the use of multi-criteria analysis for this purpose. The usefulness of DRSA is demonstrated with an illustrative example in Section \ref{sec:drsa}. Section \ref{sec:methods} then explains our methodology for collecting, preparing and processing data, and performing analysis using DRSA. The findings and analyses are then presented in Section \ref{sec:analysis}, and Section \ref{sec:conclusion} concludes.

\section{Background \label{sec:background}}
The UK government applied country-wide lockdown restrictions when the first wave of COVID-19 pandemic hit the nation in March 2020. However, as the data collection for COVID-related metrics matured, disparities within different geographic areas became more visible. It could be sub-optimal to keep the same level of restrictions to the whole country when there are rising numbers of COVID-19 positive cases (hereafter referred as 'cases' only) only in a specific local area. Hence, when facing the second wave, the UK government announced a tiered restriction system in September 2020 \citep{DepartmentOfHealthAndSocialCare}, as opposed to a country wide restrictions policy. The tiered approach may mitigate the severity of the impacts of restrictions on economic activities, such as disruptions in the critical supply chains \citep{Bassiouni2023AdvancedRestrictions}.
     
By monitoring certain measurable indicators, decisions can be made to move areas up a tier (if they are not improving) or down (if the trajectory improves). We discuss this in detail below, in order to make an argument that this is a multi-criteria decision making problem.

\subsection{The tiered system}
The tiered system, set out in the UK Government's 2020 Winter Plan \citep{Government2020}, facilitated a more systematic and data driven approach to decision making. They proposed a set of factors (criteria) to determine what level of restrictions (Tier) should be imposed on different geographical areas in England. The restriction were tiered from Tier-1 to Tier-4 \footnote{Initially only Tiers 1-3 were used and then Tier 4 came into effect later}, where Tier-1 was the most relaxed set of restriction while Tier-4 represented the most constrained level of restrictions. 

The UK government chose a set of five criteria to allocate tiered restrictions. The rationale for the choice was given as \emph{
\blockquote{The indicators have been designed to give the government a picture of what is happening with the virus in any area so that suitable action can be taken.\citep{Hancock2020}}}. The set of criteria, to determine which Tier from 1 to 4 an area should be placed in, were (C1) case detection rate in all age groups, (C2) case detection in people aged 60 or above, (C3) how quickly case rates were rising or falling, (C4) ratio of positive cases in the general population, (C5) pressure on the healthcare service. In addition to these five criteria, further consideration pertaining to the local context and exceptional circumstances could also be considered, such as a local but contained outbreak. The set of five criteria are defined and discussed below.

\subsubsection{Case detection rate in all age groups (C1)}
This criterion gives a measure of the number of cases within a given 24-hour time frame for a given geographical area. The cases were detected as positive based on the tests recorded by the UK health system. From this criteria, an indication of how many people are catching COVID-19 in an area can be gleamed. However, the measure is only based on tests that are chosen to be taken, therefore, many cases may go undetected, and differences between areas' residents reluctance to go for tests could impact the measure. 

\subsubsection{Case detection rate in the over 60 year olds (C2)}
This criterion gives a measure of the number of cases within a given 24-hour time frame (for a given geographical area) where the patient was over sixty years old. Due to the risk of serious illness from COVID-19 increasing with age this criterion is able to differentiate better the severity of the outbreak in an area than just a overall number of cases, which cannot determine if, the majority of cases are over 60 and thus more serious than if the majority of cases are under 60. Like C1, this criterion is only based on tests that are chosen to be taken and so this self selecting sample group could impact the accuracy of the measure.

\subsubsection{The rate at which cases are rising or falling (C3)}
This criterion gives a measure of the rate of change of the number of cases between one 24-hour time frame and the next. From this an indication of whether the number of cases is growing or receding can be determined. As this criterion denotes a rate of change, and not an absolute values, it can be the case that high volatility can be present when the number of cases are small. Moreover, like C1 and C2, this criterion is also based on tests that are chosen to be taken so again represents a self selecting sample group.

\subsubsection{Positivity rate (C4)}
The positivity rate criterion is a measure of the number of positive cases that are detected as a percentage of all the tests taken within a given 24-hour time frame. From this an indication of the general prevalence of COVID-19 can be extrapolated, however, it is calculated only from tests that are chosen to be taken, and potentially those that are more likely to have COVID-19 symptoms may be more likely to look to confirm this with a test. 

\subsubsection{Pressure on the NHS (C5)}

This criterion gives an indication of the pressure that the cases are having on the NHS infrastructure for a given geographical area. This is an important consideration as if the pressure were to become so high that the NHS infrastructure becomes overwhelmed then, its ability to tackle the cases that result in hospitalisation would be severally hampered, which in turn could have highly negative impacts on its ability to prevent some cases from ultimately resulting in deaths. This criterion is based on numerical information regarding hospital admissions, so unlike other criterion, such as those taking COVID-19 tests, is not based on voluntary participation. However, it only considers situations requiring health service interventions.

\subsection{Were these tiered restrictions fair?}
The government claimed that the tiered system would result in greater transparency, however, people questioned the integrity of the system from the start. For example, Manchester was forced to enter into Tier-3 without sufficient evidence to appease local policy makers and the local community. 

To investigate the issue of fairness in assigning tiered restrictions, it is important to first understand the concept of fairness. When talking of regional fairness and equality, most research has focused on the distribution of resources. For example, fair distribution of food \citep{Alkaabneh2021, Busch2016}, assigning public health resources \citep{Tavana2021, Norheim2016}, or public transport \citep{Gaile1977} between different regions. In other cases, this concept is discussed for allocating budgets (finances) in a fair and equitable manner \citep{Shankar2003}. One can argue that budget is also a type of resource distribution where the resource is a limited monetary value. Both issues involve a indirect (or indirect) competition as different regions are competing against each other for obtaining (or maximising) their share of finite resources. However, the idea of assigning different levels of restrictions does not fit into these two types, as the assignment of tier to any region is independent of how other regions are performing (or which state they are in). Thus the concept of fairness in a tiered allocation system is potentially different from the fairness in distribution of resources, where allocating more resources to one region implies the other regions will get less. Therefore, assigning heavy restrictions in one region has no dependence on the restrictions assigned to other regions. 

Although assigning restrictions to one region is independent of other regions, people have a natural tendency to compare the restrictions imposed on them with those in other regions \citep{Alicke2008}. Any preferential treatment to one region over others might not be welcomed, and might end up in some form of protest. This protest can take a form of doing (or not doing) actions that lead to breaking the rules \citep{Williams2020, Cartheyd5283}. For example, in the case of allocating tiers, people in higher tiers still have the ability to break the rules, and act as if they were in a lower tier. Therefore, when they consider it unfair, it is more likely to see people breaking the rules, which makes these decisions critical for regional peace and stability.

\subsection{North South Divide in England}
In England, the North-South divide is a term that refers to the socioeconomic differences between Southern and Northern parts of England \citep{TheEconomist2012}. A recent report by the Institute for Public Policy Research \citep{Webb2022} states that the South of England consists of one-third of the UK population yet accounts for 45\% of its economy and 42\% of its wealth. Considering these numbers, it is important to investigate the possibility of this North-South divide penetrating in the COVID-19 related policies as well. 

Before discussing this further, it is important to first define the boundaries separating the North and South. In terms of creating boundaries between North and South, a widely accepted belief is that the North consists of the regions of the North East, North West, Yorkshire and The Humber, East Midlands, and West Midlands \footnote{However, in some studies, the Midlands have also been considered a separate geographical entity from the North and the South \citep{Smith2018}) }. On the other hand, the South consists of the East of England, London, South East, and South West. Although London is considered part of South, we are interested in exploring how it has been argued that London has its own unique dynamics due to its different socioeconomic demographics. Therefore, for research purpose, we will also analyse London separately due to its unique data properties (explained in Subsection \ref{sec:eda}). 

\begin{figure}
     \centering
     \includegraphics[width=.9\textwidth]{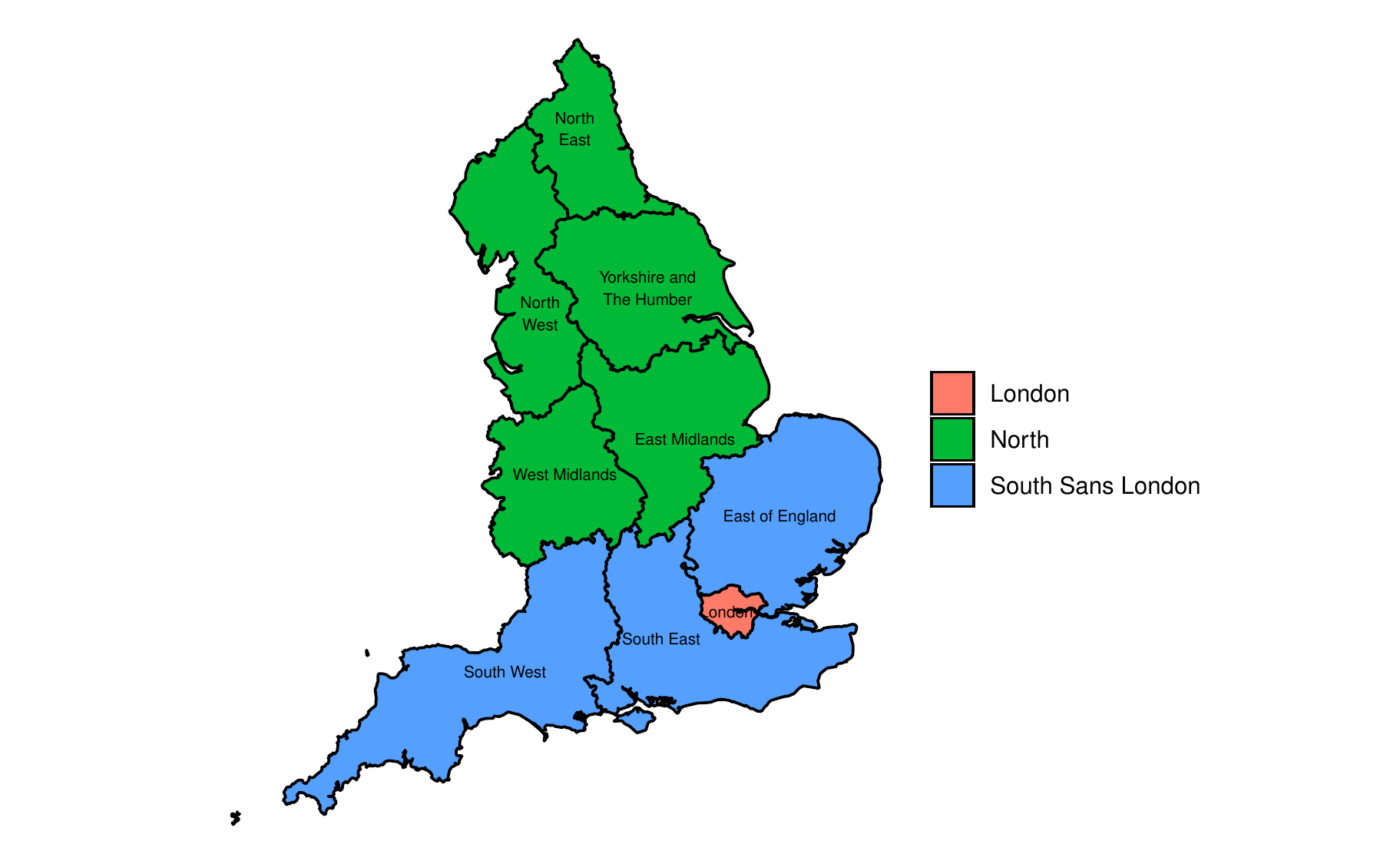}
     \caption{England Regions}
     \label{fig:northSouthDivideMap}
\end{figure}

For the purpose of our study, we will therefore consider three geographical area categories of i) North, ii) South Sans London, and iii) London. These three areas  can be seen geographically in Figure \ref{fig:northSouthDivideMap} where each of the nine regions have been coloured green, red or blue. The green regions constitute North while the blue regions constitute South without London, and London is shown in red colour.

Although the assignment of tiers is not a zero-sum game - where one person's gain is another person's loss - any inconsistency in the assignments can have serious implications as people would compare their level of freedom (or level of safety) to the other regions, and these comparisons might dissuade some of them from sticking to the rules \citep{Moss2020}. 

A recent review of the use of data analytics in COVID-19 concluded that it has been successfully employed in the healthcare sector~\citep{Khan2021ApplicationsReview}. We argue that the use of data analytical techniques can also help identify any anomalies or inconsistencies in the assignments of the tiered restrictions. In this context, the use of dominance-based rough sets approach can be useful. Before we investigate its usefulness in investigating fairness, let us briefly introduce this approach in the next section.

\section{Investigating COVID restrictions using DRSA \label{sec:drsa}}

Dominance-based Rough Set Approach (DRSA) is a well-known method for multi-criteria classification, which is used to extract understandable IF-THEN decision rules from analysing historical data \citep{Greco2001}. DRSA has been applied to numerous applications from detecting frauds \citep{Baszczynski2021AutoMethods} to analysing service quality \citep{Liou2011AQuality}. Let us demonstrate the usefulness of DRSA in analysing COVID-related data using an illustrative example below. 

Table \ref{tab:illustrative-data} provides synthetic data with ten observations, where each observation has three input readings relating to the number of cases, rate of change (in the number of cases from one day to the next), and the positivity rate. The fourth value can be considered as an output showing the Tier level allocated to that observation. In the DRSA literature, each observation is sometimes referred to as object; and the inputs are sometimes referred to as criteria attributes, and the output referred to as decision attribute.

\begin{table}[]
\centering
\caption{Illustrative example to demonstrate the use of DRSA for COVID-related data}
\begin{tabularx}{\columnwidth}{XXXXX}
\hline
    \textbf{Observation} & \textbf{Number of Cases} & \textbf{Rate of change} & \textbf{Positivity Rate} & \textbf{Tier} \\ 
\hline
$x_1$  & 195    & 2.48    & 8.05   & {3}   \\
    \rowcolor[HTML]{CCCCCC} 
$x_2$  & 92     & 2.45    & 7.89   & {2}   \\
$x_3$  & 237    & -2.74   & 8.94   & {2}   \\
    \rowcolor[HTML]{CCCCCC} 
$x_4$  & 515    & 2.82    & 1.43   & {3}   \\
$x_5$  & 528    & 7.54    & 5.3    & {3}   \\
    \rowcolor[HTML]{CCCCCC} 
$x_6$  & 434    & 1.65    & 5.41   & {2}   \\
$x_7$  & 143    & -3.15   & 8.01   & {1}   \\
    \rowcolor[HTML]{CCCCCC} 
$x_8$  & 75     & 3.2     & 5.25   & {2}   \\
$x_9$  & 269    & 2.33    & 1.71   & {1}   \\
    \rowcolor[HTML]{CCCCCC} 
$x_{10}$ & 131    & 3.28    & 1.03   & {1}  \\
\hline

 \end{tabularx}
\label{tab:illustrative-data}
\end{table}

For example, in Table \ref{tab:illustrative-data} the second observation ($x_2$) shows that there were 92 (positive) cases, while the rate of change was 2.45, and the positivity rate was  7.89\%. When comparing this observation with the first observation ($x_1$), we can see that all three input values of $x_2$ are lower than their respective values for $x_1$. Therefore, we can say that $x_2$ dominates $x_1$. In this way, we can compare the inputs of each observation with every other observation to determine its dominance. Table \ref{tab:illustrative-domination} summarises this dominance relationships for all observations. For example, the fifth observation ($x_5$) is clearly dominated by four other observations, those of $x_4$, $x_8$, $x_9$, $x_10$. Note that the table also includes $x_5$ itself as, by definition, the dominance relationship also includes equality.

\begin{table}[]
\centering
\caption{Summary of dominating and dominated observations}
\begin{tabularx}{\columnwidth}{XXX}
\hline
    \textbf{Observation} & \textbf{Dominating} & \textbf{Dominated}  \\ 
\hline
	$x_1$	&	1	&	1,2,7	\\
\rowcolor[HTML]{CCCCCC}	$x_2$	&	1, 2	&	2	\\
	$x_3$	&	3	&	3, 7	\\
\rowcolor[HTML]{CCCCCC}	$x_4$	&	4, 5	&	4	\\
	$x_5$	&	5	&	4, 5, 8, 9, 10	\\
	
\rowcolor[HTML]{CCCCCC}	$x_6$	&	6	&	6	\\
	$x_7$	&	1, 3, 7	&	7	\\
\rowcolor[HTML]{CCCCCC}	$x_8$	&	5, 8	&	8	\\
	$x_9$	&	5, 9	&	9	\\
\rowcolor[HTML]{CCCCCC}	$x_10$	&	5, 10	&	10	\\
\hline

\end{tabularx}
\label{tab:illustrative-domination}
\end{table}

After processing all the dominance relationships in Table \ref{tab:illustrative-domination}, we also need to process the outputs, which are essentially the Tier values assigned to each observation. As we know that tiers are ordered categorical values, we can group these tiers into unions by enumerating all possible combinations. This is summarised in Table \ref{tab:illustrative-unions} where we have four unions of "at most T1", "at most T2", "at least T2" and "at least T3". The unions involving "at most" are also known as \textit{downward} unions, and those involving "at least" are known as \textit{upward} unions.\footnote{Please note that the unions of "at least T1" do not make sense as all observations will be at least T1.}

\begin{table}[]
\centering
\caption{The unions of classes for the illustrative example}
\begin{tabular}{lll}
\hline
    \textbf{Union} & \textbf{Observations that are part of the union} & \textbf{Total}  \\ 
\hline
 at most T1 & {7, 9, 10} & 3 \\
\rowcolor[HTML]{CCCCCC} at most T2 & {2, 3, 6, 7, 8, 9, 10}  & 7 \\
at least T2 & {1, 2, 3, 4, 5, 6, 8} & 7 \\
\rowcolor[HTML]{CCCCCC} at least T3 & {1, 4, 5} & 3 \\

\hline

\end{tabular}
\label{tab:illustrative-unions}
\end{table}

Now that the inputs and outputs are processed separately, the next step would be to somehow link the two tables together. This is usually done by inducing rules using rough-set based algorithms like DOMLEM \citep{Greco2000} or VC-DOMLEM \citep{Baszczynski2011}. The set of rules induced using DOMLEM for this example are shown in Table \ref{tab:illustrative-rules}, where each rule has an antecedent and a consequent. The antecedent is the set of conditions that are required for this rule, and the consequent is the resulting union. 

\begin{table}[]
\centering
\caption{Illustrative example to demonstrate the use of DRSA for COVID-related data}
\begin{tabularx}{\columnwidth}{lXlll}

\hline
\textbf{No.} & \textbf{Antecedent}  &  \textbf{Consequent}  &  \textbf{Support}  &  \textbf{Strength}  \\
\hline

  1  &  \pbox{20cm}{(Number of Cases <= 269) and \\ \hspace*{1.5cm}(Positivity Rate <= 1.71)}   &  at most T1   &  2  &  66.67  \\
\rowcolor[HTML]{CCCCCC}   2  & (Rate of Change <= 2.45)  & at most T2  &  5  &  71.43  \\
  3  &  \pbox{20cm}{(Number of Cases <= 434) and \\ \hspace*{1.5cm}(Positivity Rate <= 5.41)}  &  at most T2  &  4  &  57.14  \\
\rowcolor[HTML]{CCCCCC}   4  &  (Number of Cases <= 131)  &  at most T2  &  3  &  42.86  \\
  5  &  \pbox{20cm}{(Number of Cases >= 195) and \\ \hspace*{1.5cm}(Rate of Change >= 2.48)}  &  at least T3  &  3  &  100.00  \\
\rowcolor[HTML]{CCCCCC}   6  &  (Number of Cases >= 515)  &  at least T3  &  2  &  66.67  \\
  7  &  \pbox{20cm}{(Rate of Change >= 2.82) and \\ \hspace*{1.5cm}(Positivity Rate >= 1.43)}  &  at least T2  &  3  &  42.86  \\
\rowcolor[HTML]{CCCCCC}   8  &  (Number of Cases >= 434)  &  at least T2  &  3  &  42.86  \\
  9  &  \pbox{20cm}{(Rate of Change >= 1.65) and \\ \hspace*{1.5cm}(Positivity Rate >= 5.41)}  &  at least T2   &  3  &  42.86  \\
\hline

\end{tabularx}
\label{tab:illustrative-rules}
\end{table}

Out of the nine rules extracted for this example, one can see that Rules 2, 4, 6 and 8 (see grey shaded rules) have only one condition to satisfy, and hence we can say that the rule length is 1. Consequently, the remaining five rules are of rule length 2 as all of them need two conditions to be satisfied.

The first four rules in Table \ref{tab:illustrative-rules} have \textit{consequents} involving "at most" while the remaining six rules involve "at least" in their \textit{consequents}. The rules related to "at most" are termed as downward rules, whilst the rules related to "at least" are termed as upward rules. The last two columns in Table \ref{tab:illustrative-rules} show the support and strength for each rule induced from the data. The support refers to the number of observations that adhere to this rule, while the strength refers to the ratio of observations that met the rule against the observations that only met the first part of the rule (i.e. \textit{antecedent}). For example, for the rule in row 1 of \ref{tab:illustrative-rules}, there are three observations under the \textit{antecedent}, of which 2 adhere to the rule (hence a support of 2). The strength for the rule is therefore, 3/2=66.67\%.

Some recent studies on COVID-19 demonstrate the benefits of using DRSA for data analysis and predictive modelling. For example, \citet{Figueiredo_Mota_Rosa_Souza_Lima_2022} considered a number of criteria and risk factors for COVID-19 and proposed the use of DRSA to identify areas vulnerable to COVID-19 infections in Brazil. Similarly, \citet{Bhapkar2021} proposed the use of rough sets to predict symptomatic cases of COVID-19 patients. In these and other existing studies, the focus remains on extracting a single set of rules from historical data. However, in this research, we propose a novel use of this approach, that is, to extract separate sets of rules for different geographical segments. Although these sets of rules might be useful in a number of ways, we demonstrate how we can compare these rules for investigating fairness across different geographical areas - explained in the next section.

\section{Methodology \label{sec:methods}}

The UK government's tiered allocation assignments were based on data pertaining to the prevalence and risk posed by COVID-19 in different parts of England. In the illustrative example earlier, a set of rules was derived from a set of observations. Comparisons between different geographical areas can be performed by performing this process of rules derivation for different areas of England, utilising only the observations specific for each area, to generate a separate set of rules for each area. 

In this way, from the extracted sets of rules for each geographical area under comparison, we have the opportunity to assess the overall fairness of the tiered system. To explore the fairness of the tiered approach in relation to the North-South divide, we can look to generate three separate rule sets from the observations from (i) the North, (ii) the South sans London, and (iii) London, as defined earlier and shown in Figure \ref{fig:northSouthDivideMap}.

\begin{figure}
     \centering
     \includegraphics[width=1.0\textwidth]{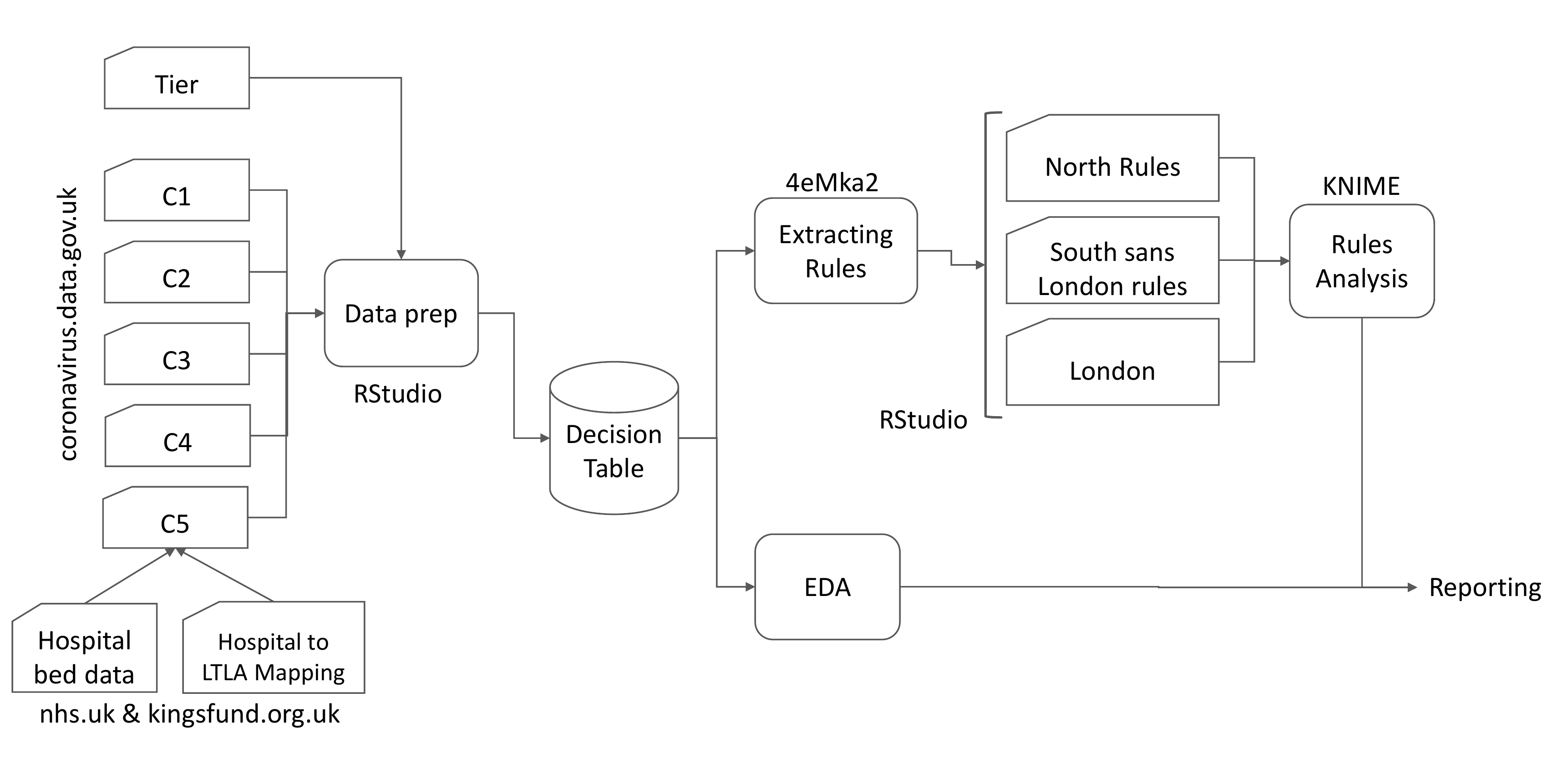}
     \caption{The Data Extraction, pre-processing and analysis pipeline of our approach}
     \label{fig:dataPipeline}
\end{figure}

The process pipeline of collecting and analysing data is shown in Figure \ref{fig:dataPipeline}; which includes data acquisition, pre-processing, and the extraction and analysis of DRSA rules. To explore such an approach requires the collection of observation data relating to the set of government criteria over the duration of the tired allocation system. For our approach data was collected from various different disparate sources. The following section explains the data collection in detail.

\subsection{Data acquisition}

For the different areas of England, data was obtained pertaining to the UK government's set of five criteria discussed earlier \citep{Hancock2020}, along with the allocated tier values. Table \ref{tab:data-sources} provides the list of data sources used to collect data on these five criteria. 

England is geographically broken up into 9 Regions which are further broken down into a set of Lower Tier Local Authority (LTLA) areas \citep{ONS2016}. Data relating to COVID-19 at the LTLA level of granularity can be collected. Therefore, we curate a data-set consisting of Tier value and criteria calculations, for each LTLA, for each day that the Tiered system was in place. We briefly outline the data acquisition process for each criterion below.

\begin{table}
    \centering
    \begin{tabular}{lll}
        \hline 
        Criterion & Description & Data source(s)  \\ 
        \hline  
        C1 & Number of Daily Cases (all ages) & \url{coronavirus.data.gov.uk} \\
        \hline  
        C2 & Number of Daily Cases in over 60s & \url{coronavirus.data.gov.uk}  \\
        \hline  
        C3 & Rate of change of daily cases &  Derived from C1 data  \\
        \hline  
        C4 & Positivity rate & \url{api.coronavirus.data.gov.uk} \\
        \hline  
        C5 & Pressure on the NHS & \makecell[l]{1) \url{api.coronavirus.data.gov.uk} \\ 2) \url{www.kingsfund.org.uk} \\ 3) \url{github.com/epiforecasts} }\\
        \hline  
        \makecell[l]{Decision \\ Variable} & Tier Value & Parliament Legislation documentation\\ 
        \hline 
    \end{tabular}
    \caption{Data sources for government-defined criteria set}
    \label{tab:data-sources}
\end{table}

\subsubsection{C1: Number of Daily Cases}

The website \url{www.coronavirus.data.gov.uk} is a UK government website that provides official data relating to coronavirus (COVID-19). The data denoting the number of positive cases which were reported in each LTLA region every day can be obtained from the site {\footnote{From the site's Supplementary downloads}}. Although the data is provided by age demographics, the overall number of cases were derived by totalling the cases across the set of age ranges. 

\begin{figure}
     \centering
     \begin{subfigure}[b]{0.49\textwidth}
         \centering
         \includegraphics[width=\textwidth]{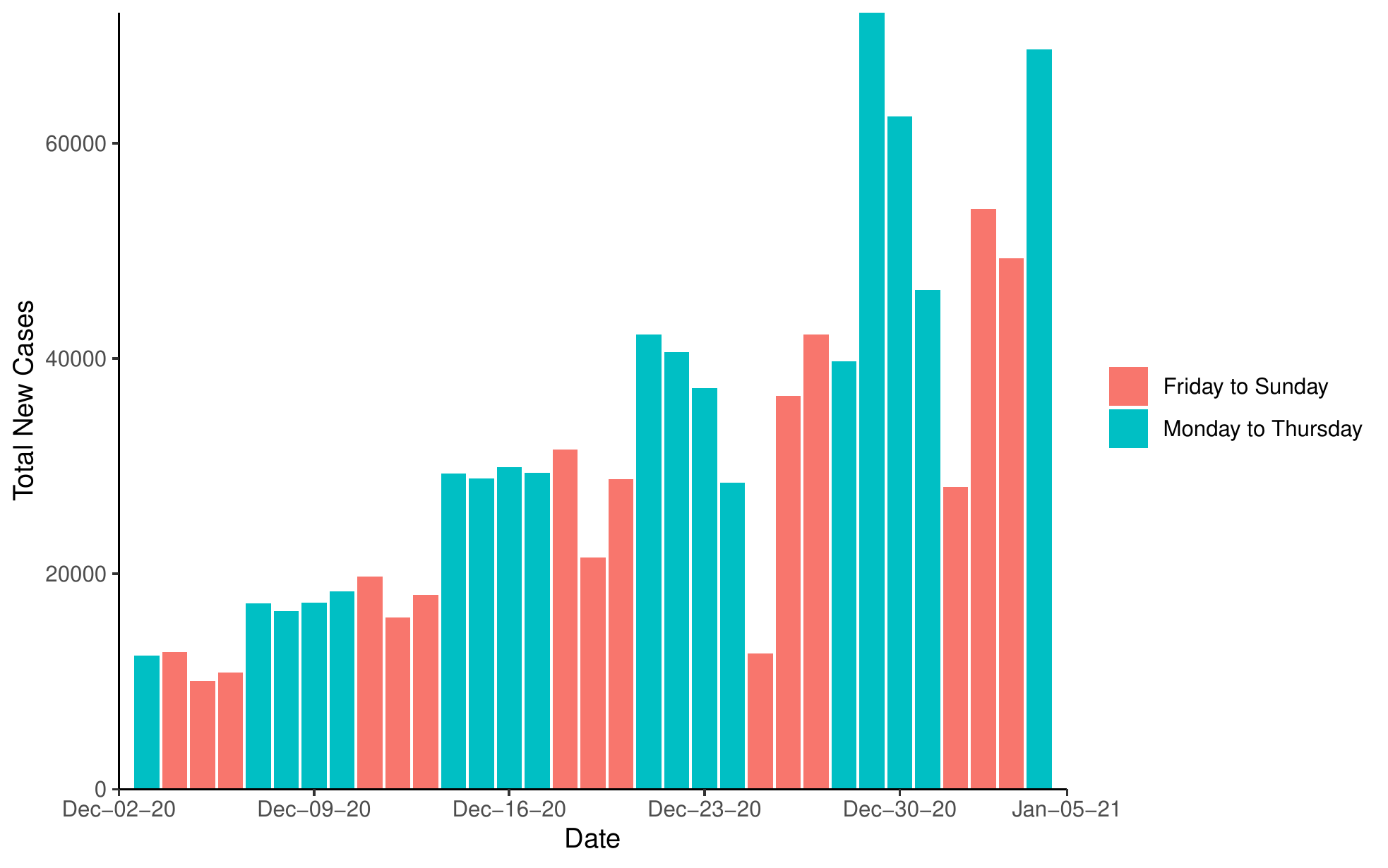}
         \caption{Raw Data}
         \label{fig:smoothingDataRaw}
     \end{subfigure}
     \hfill
     \begin{subfigure}[b]{0.49\textwidth}
         \centering
         \includegraphics[width=\textwidth]{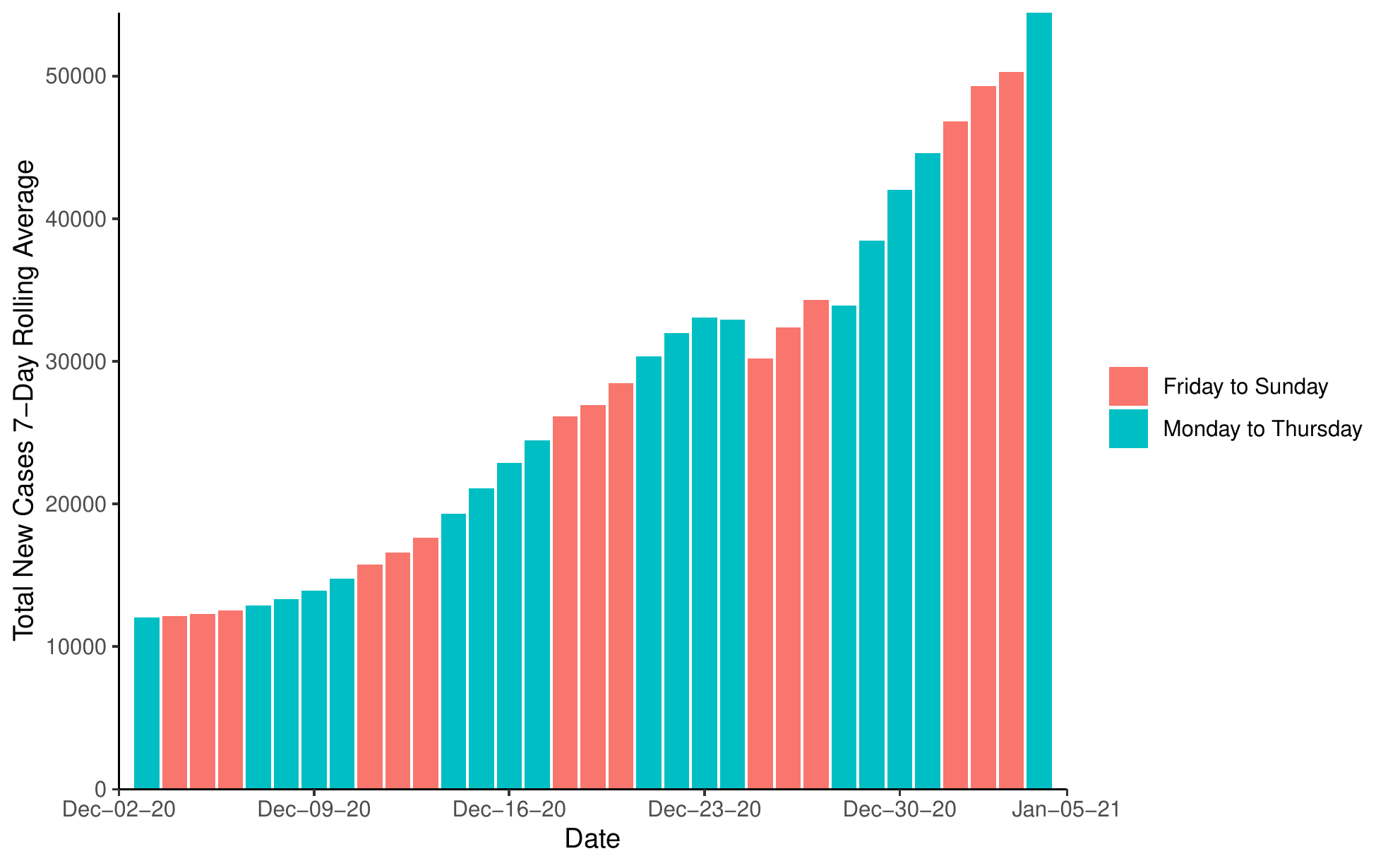}
         \caption{7-Day rolling average}
         \label{fig:smoothingDataRollingAverage}
     \end{subfigure}
     \hfill
        \caption{Raw data and 7-day rolling average data}
        \label{fig:smoothingData}
\end{figure}

As a common practice \citep{Katris2021}, we calculated a 7-day rolling average to alleviate discrepancies in the reporting velocity at different days of the week. The effects of this process are shown in Figure \ref{fig:smoothingData}, where, the figure on the left clearly shows that the number of reported cases around weekends is invariably lower than the cases on other days. The  figure on the right shows the rolling average where this issue has been addressed.

\subsubsection{C2: Number of Daily Cases in those aged 60 plus}

Recall that the the number of cases collected for C1\footnote{from the \url{coronavirus.data.gov.uk} site} are broken down by age demographic ranges, therefore, we derived C2 values as the sum of age demographic ranges 60 and over (i.e. for each LTLA for each day). As with C1, we then calculated a 7-day rolling average for C2 as well.

\subsubsection{C3:  Rate of Change in Number of Daily Cases}

The rates of change in the number of cases (from one day to the next) can be determined from the data obtained for C1. A 7-day rolling average was calculated for these rate of change values for each LTLA region for each day.

\subsubsection{C4: Positivity rate}

The positivity rate of cases denotes the percentage of tests taken over a given time period that are returning positive results. The positivity rate for each LTLA region for each day can be sought. Such data can be retrieved via the UK government's COVID-19 API \footnote{\url{api.coronavirus.data.gov.uk} - providing API calls for a range of COVID-19 related metrics and levels of geographical granularity.} Through utilising the API making a set of call requests, for each LTLA region, we obtained the positivity rate, for each LTLA region for each day, which again we processed into 7-day rolling average values.

\subsubsection{C5: Pressure on the National Health Service (NHS)}
    
The criterion of pressure on the NHS looks to measure the amount of pressure being placed on the local medical services, in terms of their ability to be able to cope with the COVID-19 cases prevalent within a given area. The UK government utilised this criterion in their tiers allocation system, however, the pressure on NHS cannot be quantified directly at LTLA level. To estimate the pressure, we relied on the number of beds occupied by COVID-19 patients at NHS Trusts level. The NHS service is broken geographically across England into a number of Trusts, 223 in total\footnote{\url{www.kingsfund.org.uk/audio-video/key-facts-figures-nhs}} (which each operate as a somewhat automated unit). 

Due to lack of government transparency in quantifying the pressure on NHS, the Trust level values were collected and mapped to LTLA level to quantify the pressure on the NHS, as there was a need to align the Trust level data with the LTLA level data. This mapping (performed in R) is explained below.

Using the UK government's COVID-19 API provides the number of beds occupied by COVID-19 cases (for each trust for a given day). NHS trusts across England vary significantly in size (and thus levels of capacity) therefore, comparisons between such values on their own may be misleading. For example, one trust may have double the number of beds occupied with COVID-19 patients than another but still be under less pressure due to being three times the size. The government documentation regarding the tiers system remains opaque in terms of how such information was exactly utilised to help make decisions. However, we can utilise various other publicly available data to derive measures of the relative pressure on the NHS.

Each trust has multiple hospital sites for which data relating to the total number of available beds for each trust can be sought {\footnote{\url{www.kingsfund.org.uk}}}. We already know the actual number of beds occupied by COVID-19 patients so therefore, we can estimate a representative ratio of occupied beds in relation to the overall number of beds in each trust. 

Recall these values are available for each trust, however, to perform these analysis, we need all values at the LTLA level. A trust may serve more than one LTLA region, and an LTLA region might be served by multiple trusts. Therefore, in this way, there exists a many to many mapping between trusts and LTLAs. 

One way to resolve this issue is to calculate probabilistic estimates through mapping the information at trust level to LTLA level. To achieve this we utilised the Trust to local authority mapping provided by Epiforecasts\footnote{Available as a developer package on GitHub at - \url{ github.com/epiforecasts/covid19.nhs.data}}. This package provides probabilistic mapping estimates that can be used to estimate COVID-19 hospital admissions at an LTLA level.
From utilising this mapping on our data we obtained a ratio based pressure value for each LTLA region for every day, and thus at the same level of granularity that we have for the other four criteria.

\subsubsection{Decision Variable: Tiers Data}
Date pertaining to tier values (1,2,3 or 4) that different parts of England were placed under, and subsequent changes, were announced in the houses of Parliament and published in official legislation documentation by the government \citep{PublicHealth2020}. From this we were able to define the Tier that each LTLA region was in for every day that the tired system was in place. 


\subsection{Data processing and integration}

From the acquisition and format alignment processing of data from these various sources making up the set of 5 criteria we then integrated the data to create an overall data-set. In the data-set each row (observation) denotes the data for the set of 5 criteria and the tier pertaining to an LTLA and Date pair, with additional information added to each observation regarding its membership of either Northern England, Southern England sans London, or London. The data-set consists of 10827 observations. \footnote{The enriched data-set for this study is publicly available on github here: \href{https://github.com/prioritization/DRSA\_Covid19}{https://github.com/prioritization/DRSA\_Covid19}}

The data-set was then utilised to gain understanding about the allocation of tiers,  first with an exploratory data analysis, and then using the DRSA rules analysis approach. These two approaches are described in the next subsections below.

\subsection{Exploratory Data Analysis}
For better data understanding, we performed some statistical analysis of the data-set. This included analysing the spread of decision variable values, across the range of dates, and between the different geographical areas. We then analysed the relationships between and within the criteria set and tier values allocated. The results for exploratory data analysis are discussed in Section \ref{sec:eda}.

\subsection{DRSA Rule Extraction}
 
To explore the fairness and validity of the tiers allocation system via DRSA, rule extraction and comparison of rule sets was performed.
 
The overall data-set was divided into three separate files representing observations pertaining to the North, the South sans London and London. These  observation files were then processed \footnote{using R scripts to automate the process} to generate information system files (.isf) readable by 4eMka2\footnote{4eMka2 is a tool developed by the Laboratory of Intelligent Decision that implements the DRSA method Support System of the Institute of Computing science, Poznan University of Technology, \href{idss.cs.put.poznan.pl}{idss.cs.put.poznan.pl}}. These  files were then fed into 4emka2 that generated three output files, providing us the three separate sets of extracted rules files (.rls). This process is also summarised in Figure \ref{fig:dataPipeline} (see top-right of the figure). Regarding rule extraction parameters, initially only the rules with a minimum rule strength of 25\% were saved, with a maximum rule length of up to 5 (i.e. all of the criteria). The output .rls files were then parsed and tabulated (using R scripts for automation) for further processing. Then, using the mining tool KNIME \footnote{An open-source data analytics, reporting and integration platform - \href{www.knime.org}{www.knime.org}} the separate rules were analysed and compared.

\section{Analysis and Results\label{sec:analysis} }
   
In this section the results of the exploratory data analysis, and the analysis and discussions of using the DRSA rules analysis approach are described.


\subsection{Exploratory data analysis \label{sec:eda}}

A set of exploratory data analyses were carried out to gain data understanding before investigating further. These are described and discussed below.

\subsubsection{Spread of tiers values}
\begin{figure}
     \centering
     \begin{subfigure}[b]{0.49\textwidth}
         \centering
         \includegraphics[width=\textwidth]{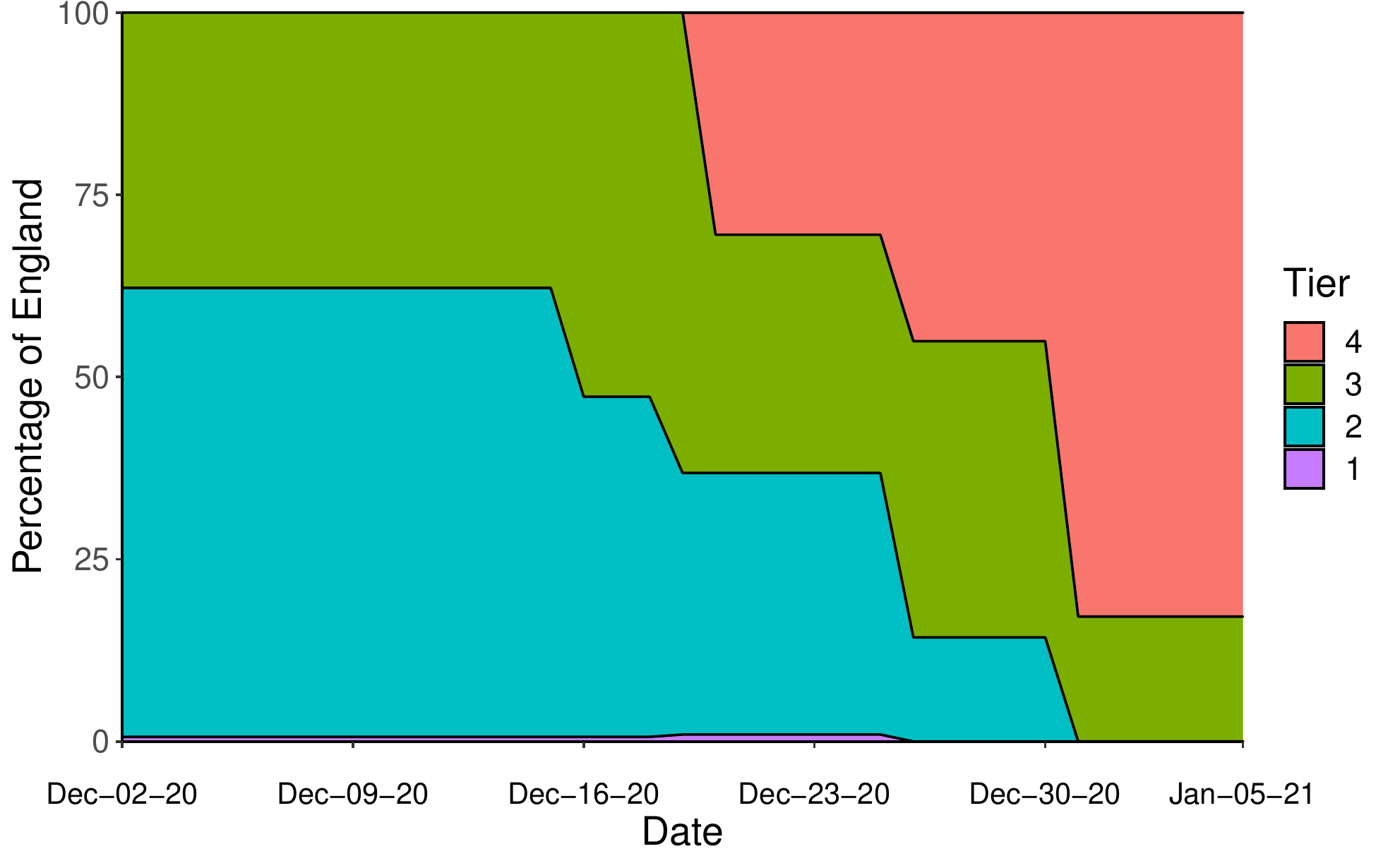}
         \caption{ALL}
         \label{fig:areaPlotAll}
     \end{subfigure}
     \hfill
     \begin{subfigure}[b]{0.49\textwidth}
         \centering
         \includegraphics[width=\textwidth]{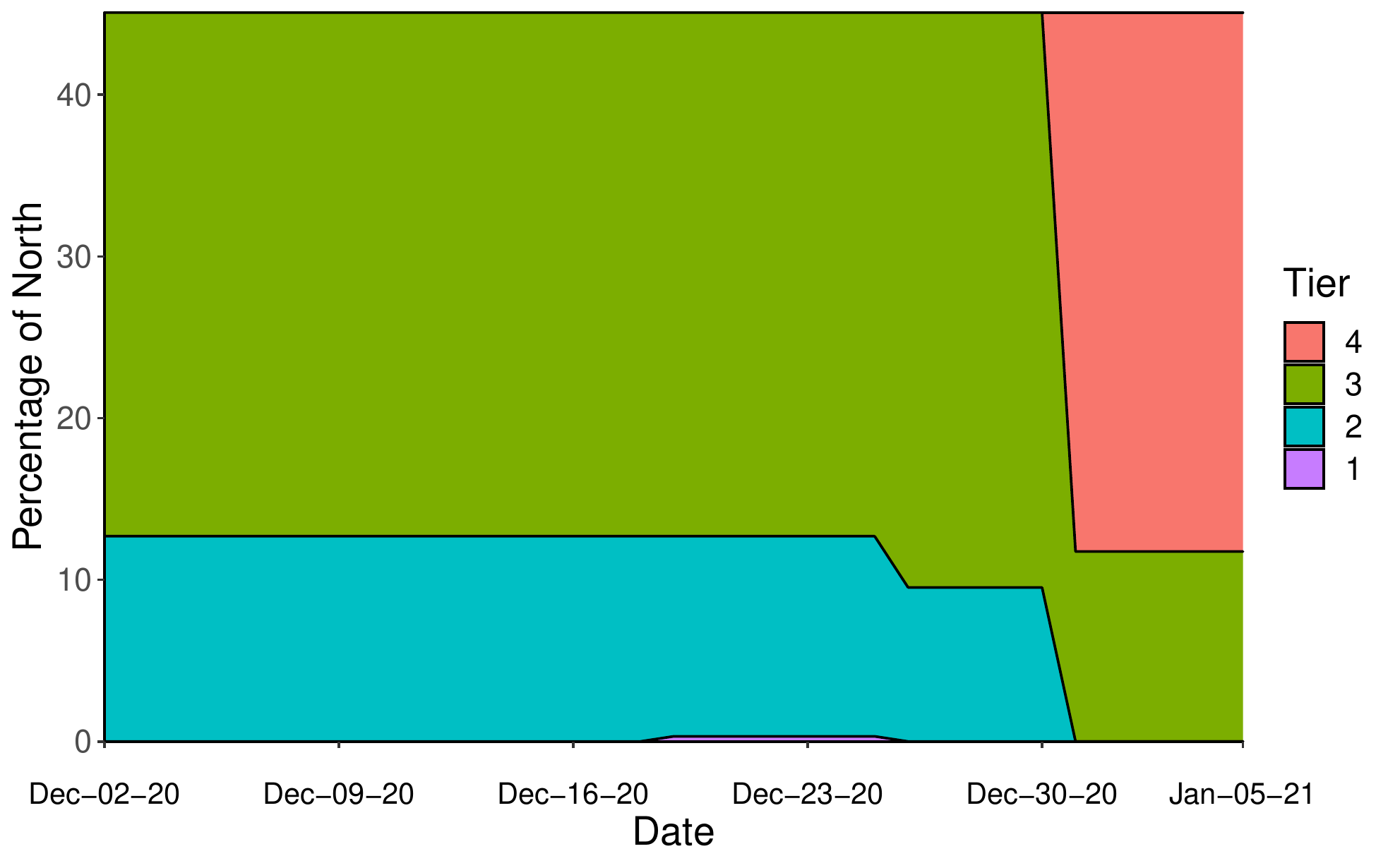}
         \caption{North}
         \label{fig:areaPlotNorth}
     \end{subfigure}
     \hfill
     \begin{subfigure}[b]{0.49\textwidth}
         \centering
         \includegraphics[width=\textwidth]{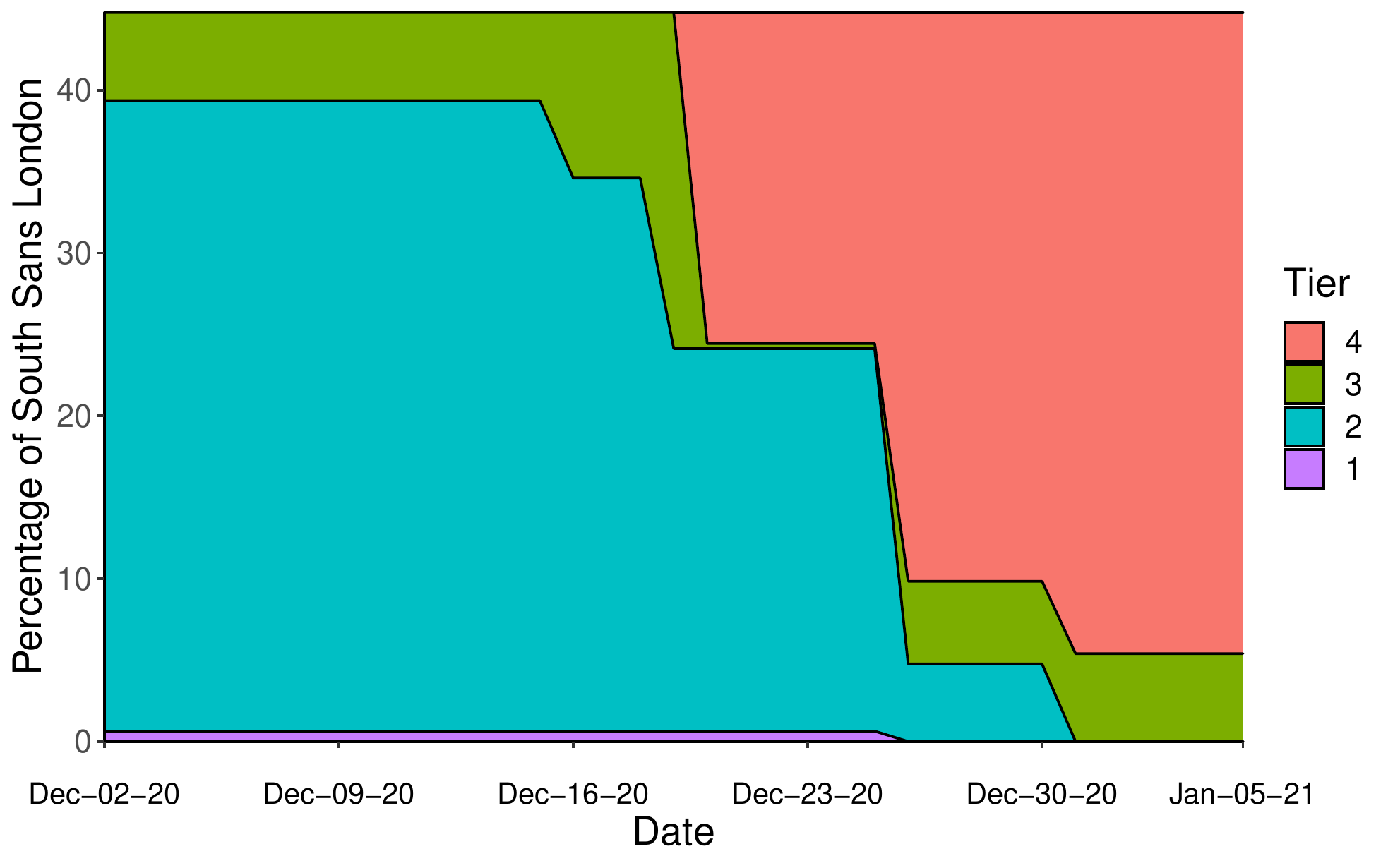}
         \caption{South sans London}
         \label{fig:areaPlotSouth}
     \end{subfigure}
     \hfill
     \begin{subfigure}[b]{0.49\textwidth}
         \centering
         \includegraphics[width=\textwidth]{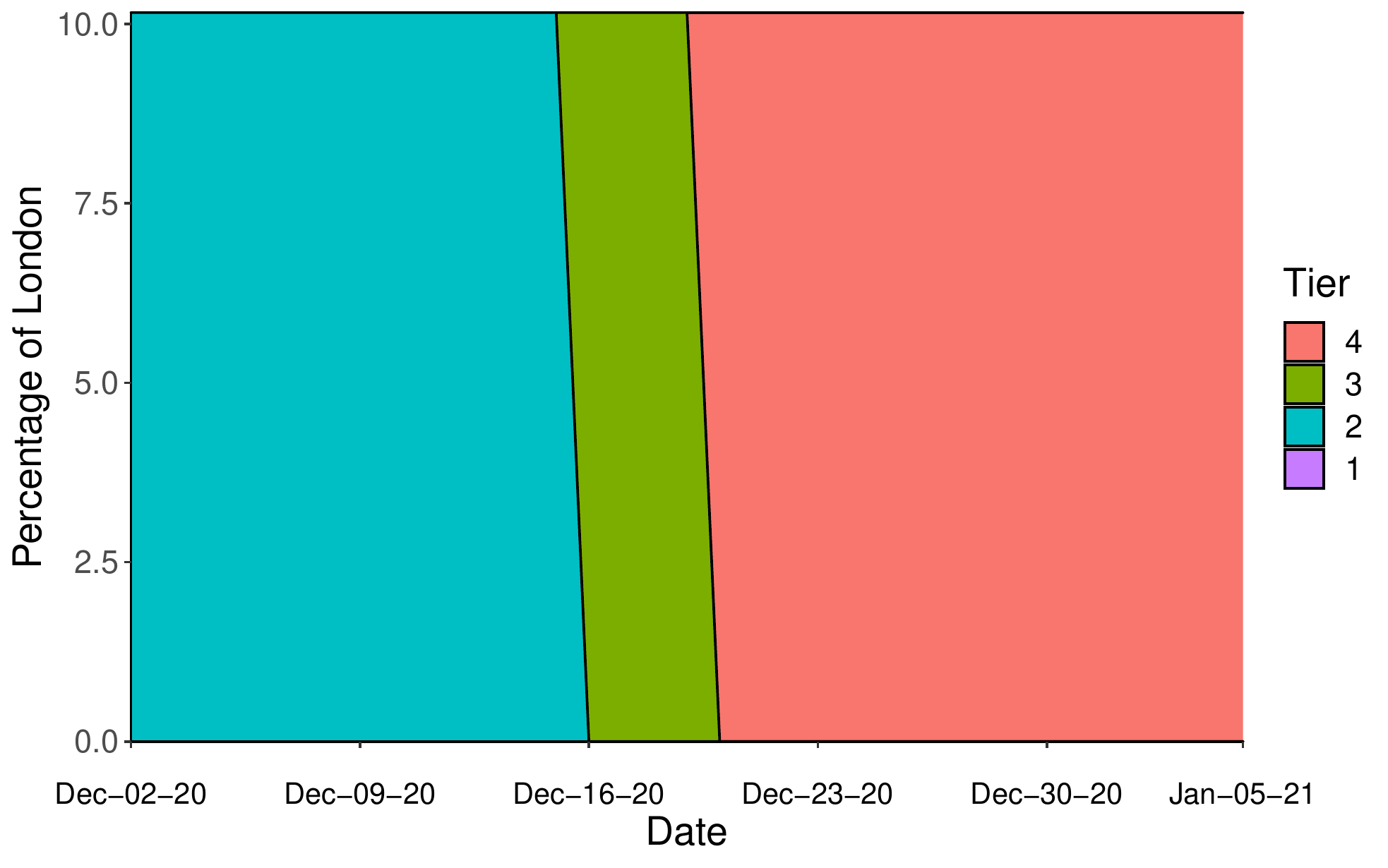}
         \caption{London}
         \label{fig:areaPlotLondon}
     \end{subfigure}
     \hfill
        \caption{Density Plots of the Distribution of Tiers across the Data Range for different geographic areas}
        \label{fig:areaPlot}
\end{figure}

We explored the distribution within the data of Tier values across the time period. Figure \ref{fig:areaPlotAll} shows these distributions for the whole data-set, where it can be seen that a majority of areas were in Tier-2 in early November but moved into Tier-4 by mid December. Tier-1, however, remains infrequent in the whole data-set, with only 0.5\% of observations, (the other tier values percentages of observations in the whole data-set are Tier-2, 37.9\%, Tier-3, 37.8\%, and Tier-4, 25.9\%). 

Figure \ref{fig:areaPlotNorth} shows that the majority of the Northern regions started and stayed in Tier-3 while the Southern regions (sans London) started in Tier-2 (see \ref{fig:areaPlotSouth}) and stayed in Tier-2 until towards the end of December. London followed a similar but even more pronounced pattern than the South, as visible in \ref{fig:areaPlotLondon}. 

The analysis of Figure \ref{fig:areaPlot} highlights the imbalance of the distribution of the Tier values across different regions. This imbalance might impact the coverage of DRSA rules that can be generated, for example, rules about Tier-1 might not be easy to derive due to the sparsity of observation data with Tier-1.


\begin{theorem}There is an imbalance of the distribution of the Tier values.\end{theorem}


\subsubsection{Analysis of relationships between Tier values and criteria values}

Next, we analyse potential correlations between the criteria values and Tier values, across the different geographical areas. These correlations are shown in a set of tables provided in Table \ref{tbl:correlationTables}. Looking at Table \ref{tbl:correlationTables}a - showing pairwise correlations among criteria and Tiers within the whole data-set - one can see that there is a strong positive correlation between C1 and C2 (i.e. $0.94$). Although this is true for all four tables, the correlation in the data from North (see Table  \ref{tbl:correlationTables}b) is slightly stronger than in the data obtained from The South Sans London and London (see Tables  \ref{tbl:correlationTables}c and  \ref{tbl:correlationTables}d). However, regarding the other criteria the North has significantly lower correlations amongst the set of five criteria (and Tier values)
than The South Sans London and London.

\begin{theorem}The number of cases in age group 60+ (C2) is highly correlated with the overall number of cases (C1)\end{theorem} 
\begin{theorem}Data for the North has significantly lower correlations amongst the set of five criteria (and Tier values)\end{theorem} 

A further insight from the correlation data is that C3 (Total New Cases Rate of Change) has a weak correlation with the Tier value (as well as to the other criteria). Conversely, C4 (Positively Rate) has the strongest correlation with the Tier value for the overall data (as well as for each of the three geographical breakdowns). This suggests that the 'Total New Cases Rate of Change' has little influence upon Tier values, whilst the 'Positivity Rate' does. 

\begin{table}[!ht]
	\begin{tabular}{cc}
	    \begin{minipage}{.5\linewidth}
            \begin{tabular}{l|rrrrrr}
                 & C1   & C2    & C3    & C4    & C5    & Tier  \\ \hline
            C1   &  & 0.94  & 0.01  & 0.74  & 0.61  & 0.49  \\
            C2   &  &  & -0.02 & 0.74  & 0.54  & 0.53  \\
            C3   &  &  &  & -0.06 & -0.06 & -0.10 \\
            C4   &  &  &  &  & 0.56  & 0.71  \\
            C5   &  &  &  &  &  & 0.36   \\
            Tier
            \end{tabular}
            \label{tbl:correlationTableALL}
            \\ \centering{ (a) Correlations for overall data}
	    \end{minipage} &
	
	    \begin{minipage}{.5\linewidth}
            \begin{tabular}{l|rrrrrr}
                 & C1   & C2    & C3    & C4    & C5    & Tier  \\ \hline
            C1   &  & 0.96  & 0.12  & 0.52 & 0.26  & 0.32  \\
            C2   &  &  & 0.11 & 0.53  & 0.25  & 0.34  \\
            C3   &  &  &  & 0.11 & -0.03 & -0.00 \\
            C4   &  &  &  &  & 0.25  & 0.65  \\
            C5   &  &   &  &  &  & 0.28   \\
            Tier
            \end{tabular}
            \\ \centering{ (b) Correlations for North region}            
	    \end{minipage} 
	\end{tabular}
	
	\bigskip
	\bigskip
		
	\begin{tabular}{cc}
	    \begin{minipage}{.5\linewidth}
            \begin{tabular}{l|rrrrrr}
                 & C1   & C2    & C3    & C4    & C5    & Tier  \\ \hline
            C1   &  & 0.93  & -0.09  & 0.80  & 0.68  & 0.58  \\
            C2   &  &  & -0.13 & 0.80  & 0.62  & 0.62  \\
            C3   &  &  &  & -0.18 & -0.10 & -0.16 \\
            C4   &  &  &  &  & 0.61  & 0.75  \\
            C5   &  &  &  &  &  & 0.40   \\
            Tier
            \end{tabular}
            \\ \centering{ (c) Correlations for South region Sans London}	    \end{minipage} &
	
	    \begin{minipage}{.5\linewidth}
            \begin{tabular}{l|rrrrrr}
                 & C1   & C2    & C3    & C4    & C5    & Tier  \\ \hline
            C1   &  & 0.93  & -0.27  & 0.78  & 0.41  & 0.62  \\
            C2   &  &  & -0.10 & 0.76  & 0.44  & 0.64  \\
            C3   &  &  &  & -0.16 & -0.11 & -0.12 \\
            C4   &  &  &  &  & 0.52  & 0.76  \\
            C5   &  &   &  &  &  & 0.45   \\
            Tier
            \end{tabular}
            \\ \centering{ (d) Correlations for Greater London}       
	    \end{minipage}

	\end{tabular}

\caption{Correlation table...}
\label{tbl:correlationTables}
\end{table}

\begin{figure}[!ht]
     \centering
     \begin{subfigure}[b]{0.49\textwidth}
         \centering
         \includegraphics[width=\textwidth]{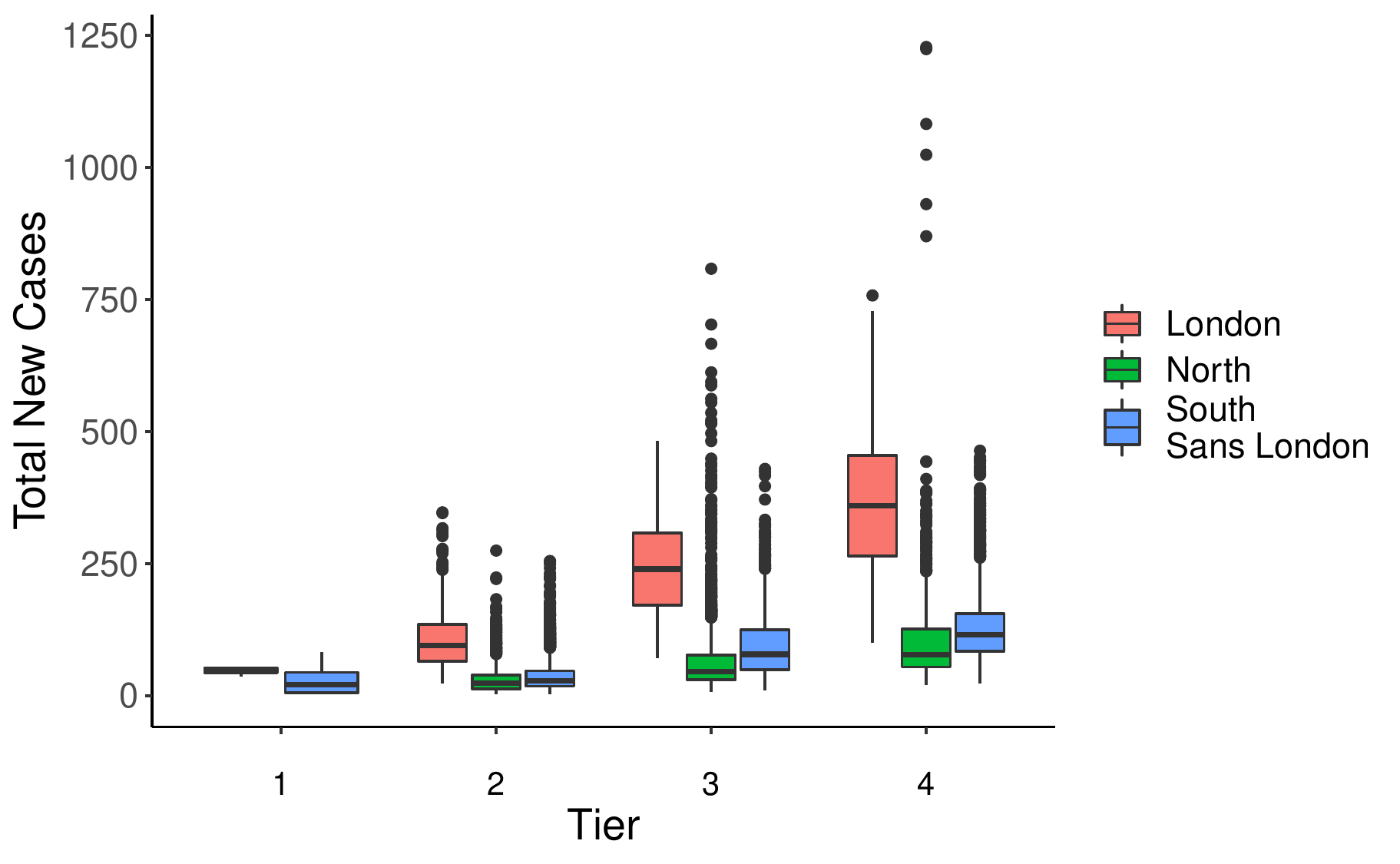}
         \caption{C1 - Total New Cases}
         \label{fig:boxPlotC1}
     \end{subfigure}
     \hfill
     \begin{subfigure}[b]{0.49\textwidth}
         \centering
         \includegraphics[width=\textwidth]{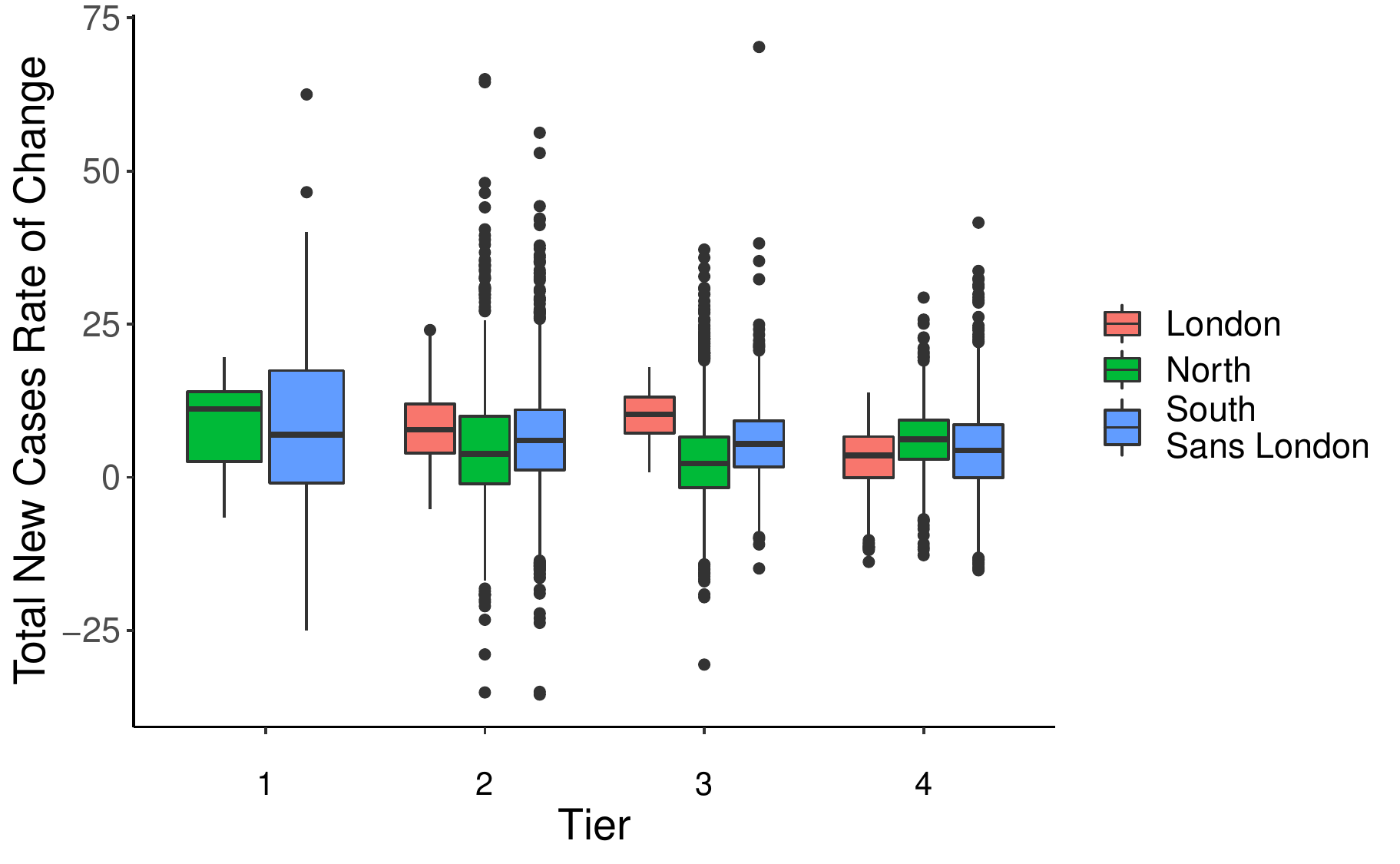}
         \caption{C3 - Total New Cases Rate of Change}
         \label{fig:boxPlotC3}
     \end{subfigure}
     \hfill
     \begin{subfigure}[b]{0.49\textwidth}
         \centering
         \includegraphics[width=\textwidth]{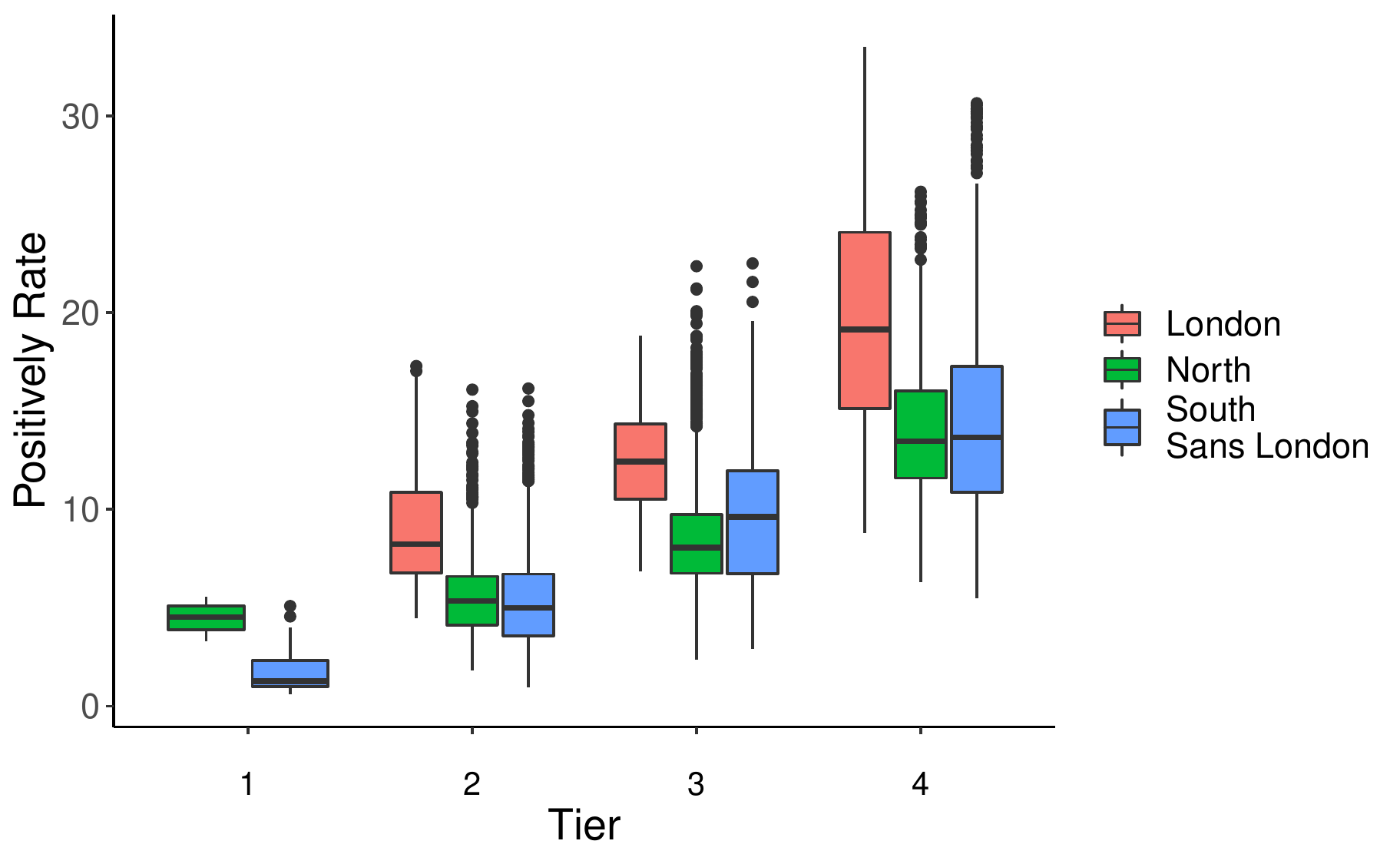}
         \caption{C4 - Positivity Rate}
         \label{fig:boxPlotC4}
     \end{subfigure}
     \hfill
     \begin{subfigure}[b]{0.49\textwidth}
         \centering
         \includegraphics[width=\textwidth]{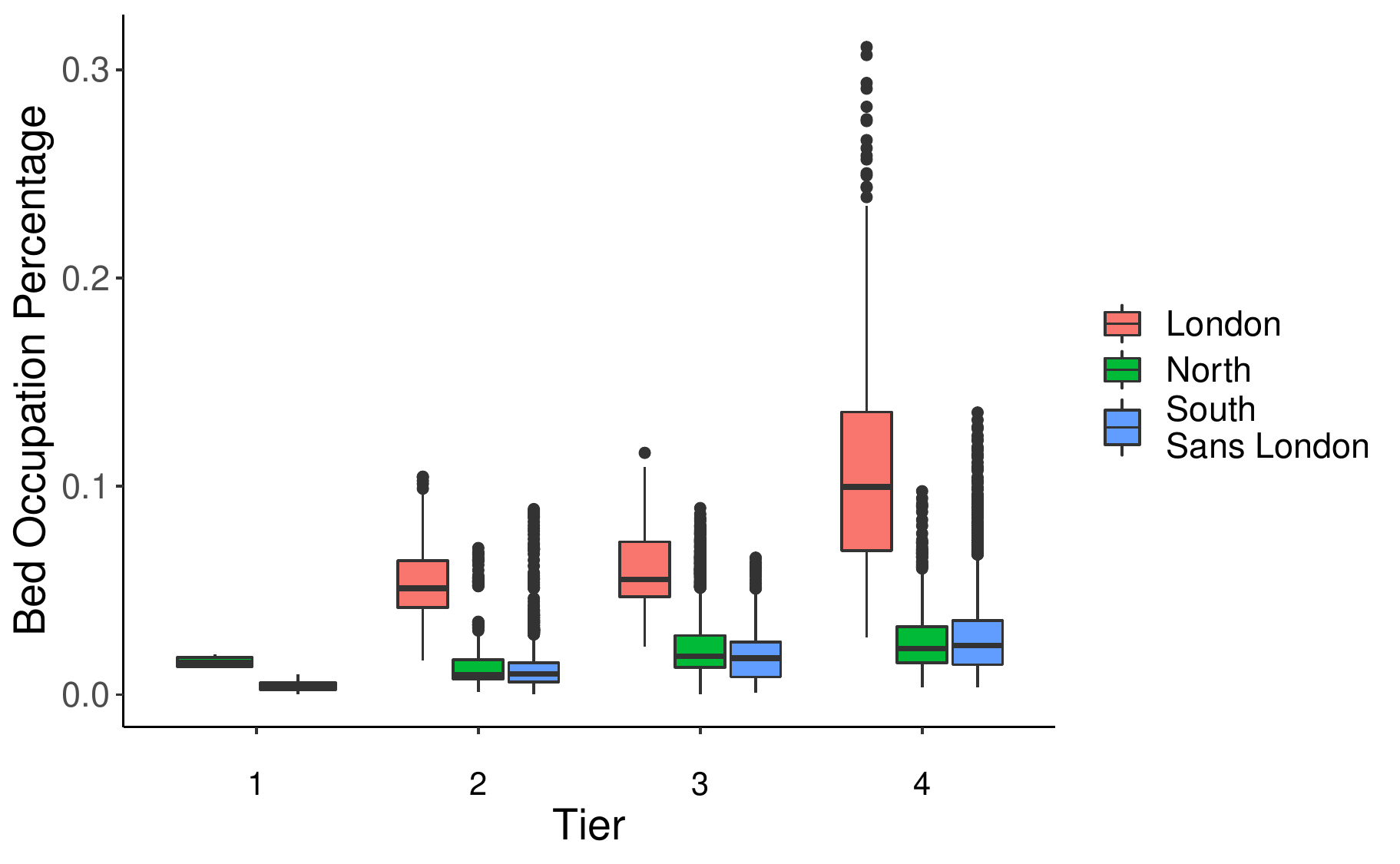}
         \caption{C5 - Pressure on the NHS}
         \label{fig:boxPlotC5}
     \end{subfigure}
     \hfill
        \caption{Box Plots}
        \label{fig:boxPlots}
\end{figure}

These relationships are shown visually through Box-plots in Figure \ref{fig:boxPlots}. Here, Box-plots are shown for C1, C3, C4 and C5. C2 is omitted due to the strong positive correlation previously identified between C1 and C2, which results in very similar box-plots for C1 and C2. In each of these plots, the criteria values are grouped according to the Tier values, and a statistical summary is shown for North, South Sans London, and London. For C1, C4 and C5, the Box Plots show that London was invariably assigned to be in a lower Tier for values that would have resulted in higher Tier assignment for the North and the South Sans London. For example, for C1, the median value for London being assigned Tier-3 is 239.43, where as conversely, the median value for North and South being assigned to Tier-4 is 77.14, and 115.50000 respectively. This suggests London was given different treatment, perhaps showing that is was being favoured for economic activities at the expense of public health; suggesting implicit additional criteria exist within the Tiers system.

\begin{theorem}For C1, C4 and C5, London was allocated to be in Tier-2 or Tier-3 for the values that put other geographical regions in Tier-4.\end{theorem}

From these box-plots, for C3 (Total New Cases Rate of Change) shown in Figure \ref{fig:boxPlotC3}, there is apparently no relation between the criterion's value and the assigned Tier. Whereas, for C4 (Positivity Rate), a clear positive relationship is visible with the Tier value (albeit at different rates for the different geographical regions). This reiterates what the correlation analysis highlighted, namely that C3 appears to have a weak or no relationship with an assigned Tier value where as C4 does. 

\begin{theorem}The Rate of change (C3) has a weak or no relation with Tiers allocation, whilst C4 (Positivity Rate) appears to have a strong relation with Tiers.\end{theorem} 

So far, each criterion has been investigated individually, without analysing possible interrelationships amongst them and the assigned Tiers. Also, to investigate the fairness, it is more appropriate to investigate the combinations of tiers, for example, investigating geographical area in "at least" Tier-3 or "at most" Tier-2. Such analyses can be carried out using DRSA, as discussed in the next subsection. The DRSA models generated from such analysis can also be used to help us sort unseen data, and in turn, facilitate predictive analysis for evidence-based decision making. 

\subsection{DRSA Rule analysis and comparison}
Although the DRSA has been widely used in practical applications, these applications have focused on a single set of rules extracted from a data-set. On the contrary, in this research, we focus on generating multiple sets of rules created from different data-sets, and then comparing these rule sets for gaining further insights. 

\begin{figure}[t!]
     \centering
     \includegraphics[width=0.8\textwidth]{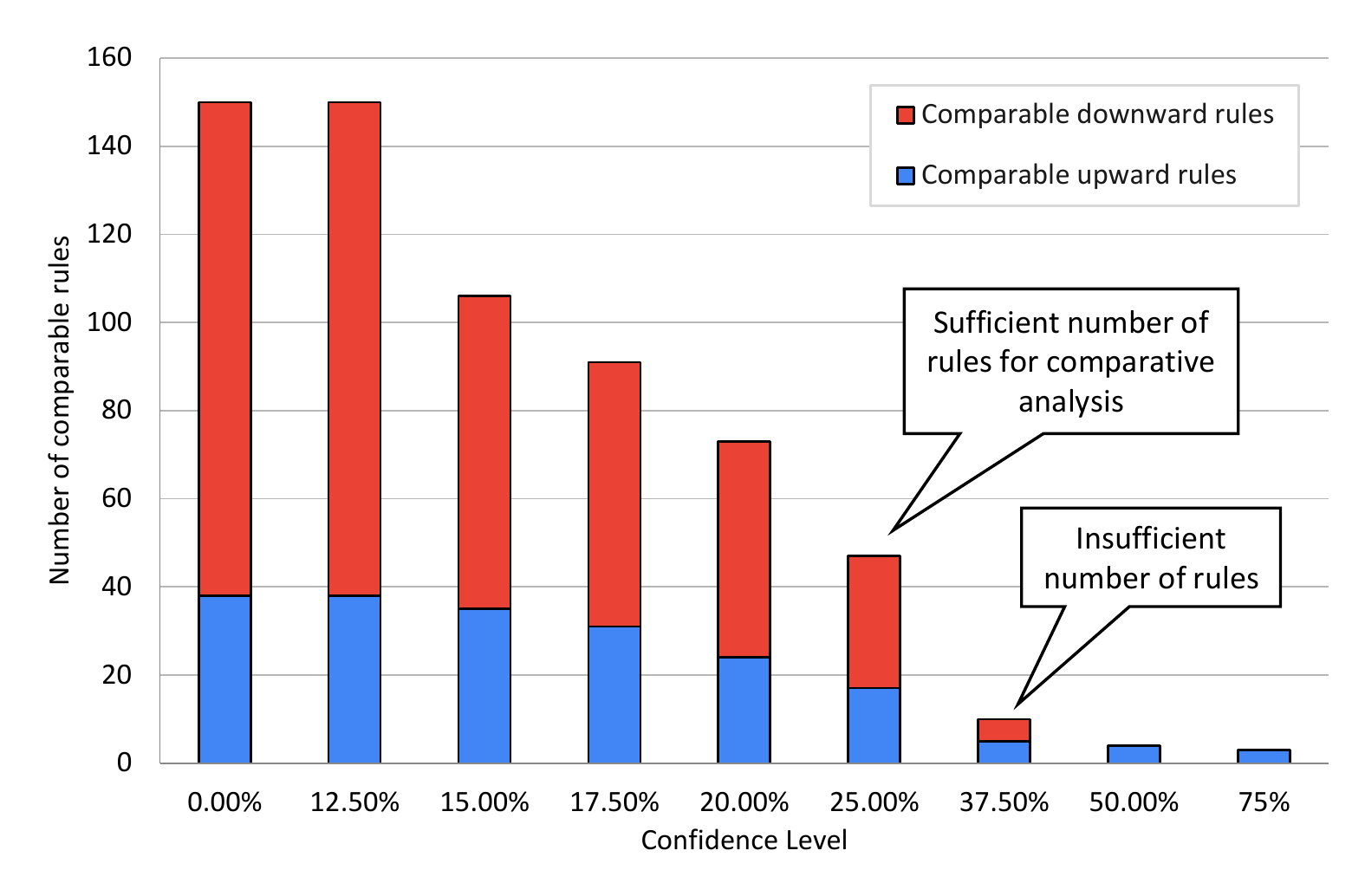}
     \caption{Increasing the confidence threshold reduces the number of possible comparisons}
     \label{fig:confidenceThresholds}
\end{figure}

The rules were extracted using the 4emka software tool using the rule strength as a filter to include/exclude rules for comparison. Keeping the rules with lower strength may reduce the quality of information for comparing different geographical regions. However, on the other hand, filtering out these rules might result in having sparse information which could be insufficient for comparative analysis. This can be seen as a quantity-quality trade-off problem and finding a balance between these two objectives is an important issue to address. Recall the lack of coverage of our data-set shown in Figure \ref{fig:areaPlotAll}. This already constrains our analysis due to a limited number of rules having higher confidence values. 

Figure \ref{fig:confidenceThresholds} shows the results of an experimental trade-off analysis carried out by changing the threshold level of confidence which can help determine a trade-off between confidence level (i.e. quality) and the number of comparable rules (i.e. quantity). As shown in this figure, the number of comparable rules decrease significantly as we increase the threshold of acceptance for confidence. Therefore, in this case study, we used the minimum rule strength of 25\% to keep sufficient information required to compare different regions. However, we believe the selection of this threshold is both 1) context-dependent, where other input data sets might result in alternative thresholds to be chosen, and 2) User dependent, subjective to different user dispositions regarding the trade-offs involved.

\begin{table}[t!]
     \centering
     \includegraphics[width=1\textwidth]{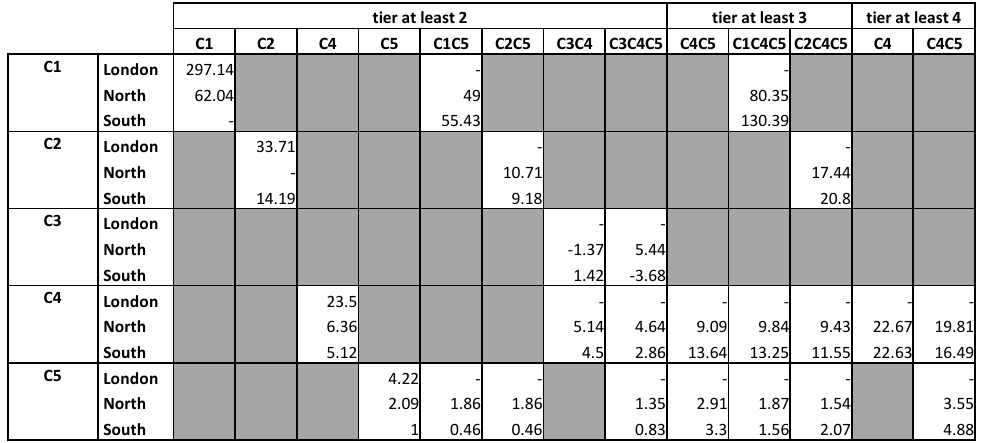}
     \caption{Upwards Rules Table}
     \label{tab:UpwardRules}
\end{table}

\begin{figure}
     \centering
     \includegraphics[width=1\textwidth]{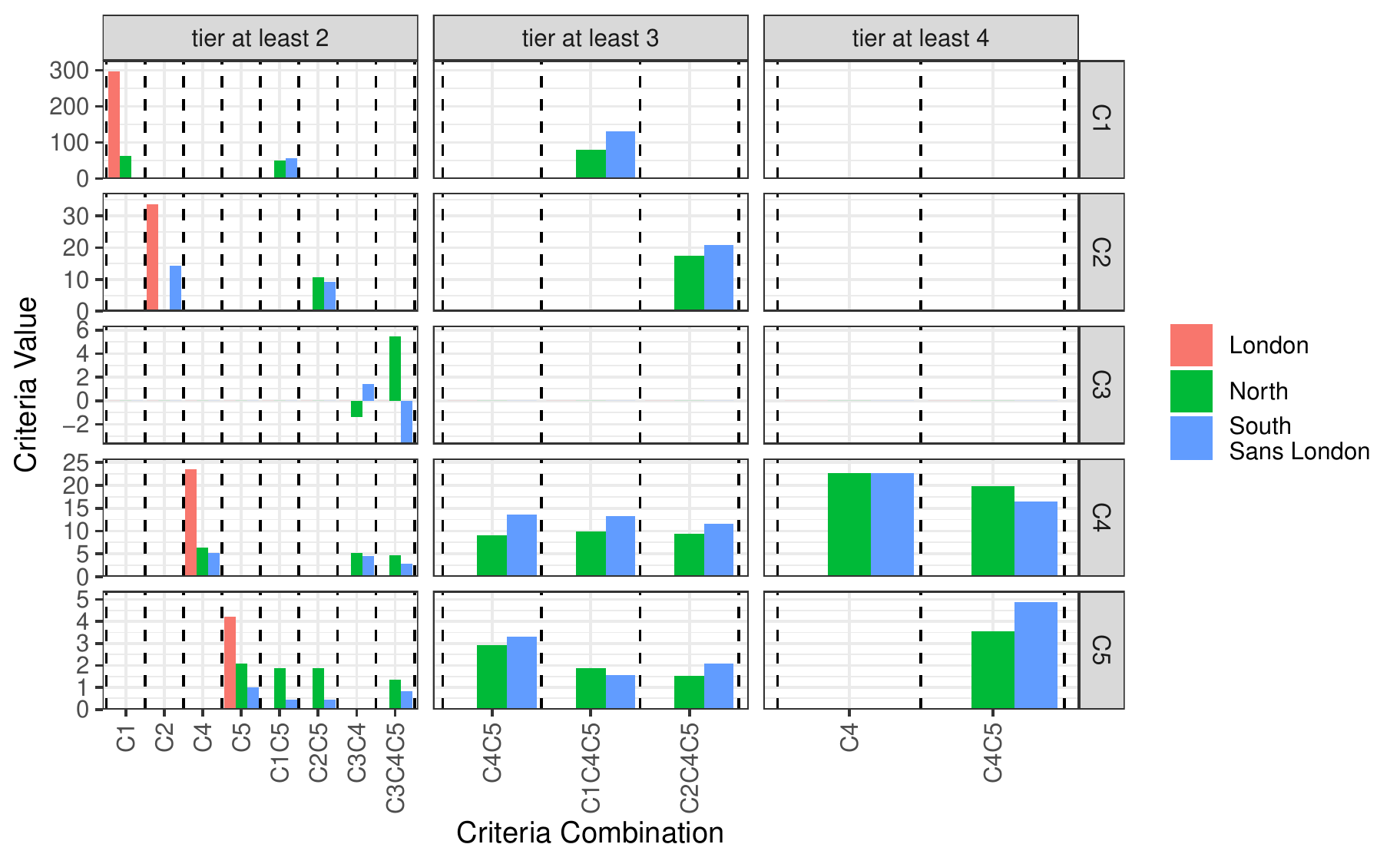}
     \caption{Upward Rules Plot}
     \label{fig:UpwardRules}
\end{figure}

\begin{figure}[t!]
     \centering
     \includegraphics[width=1\textwidth]{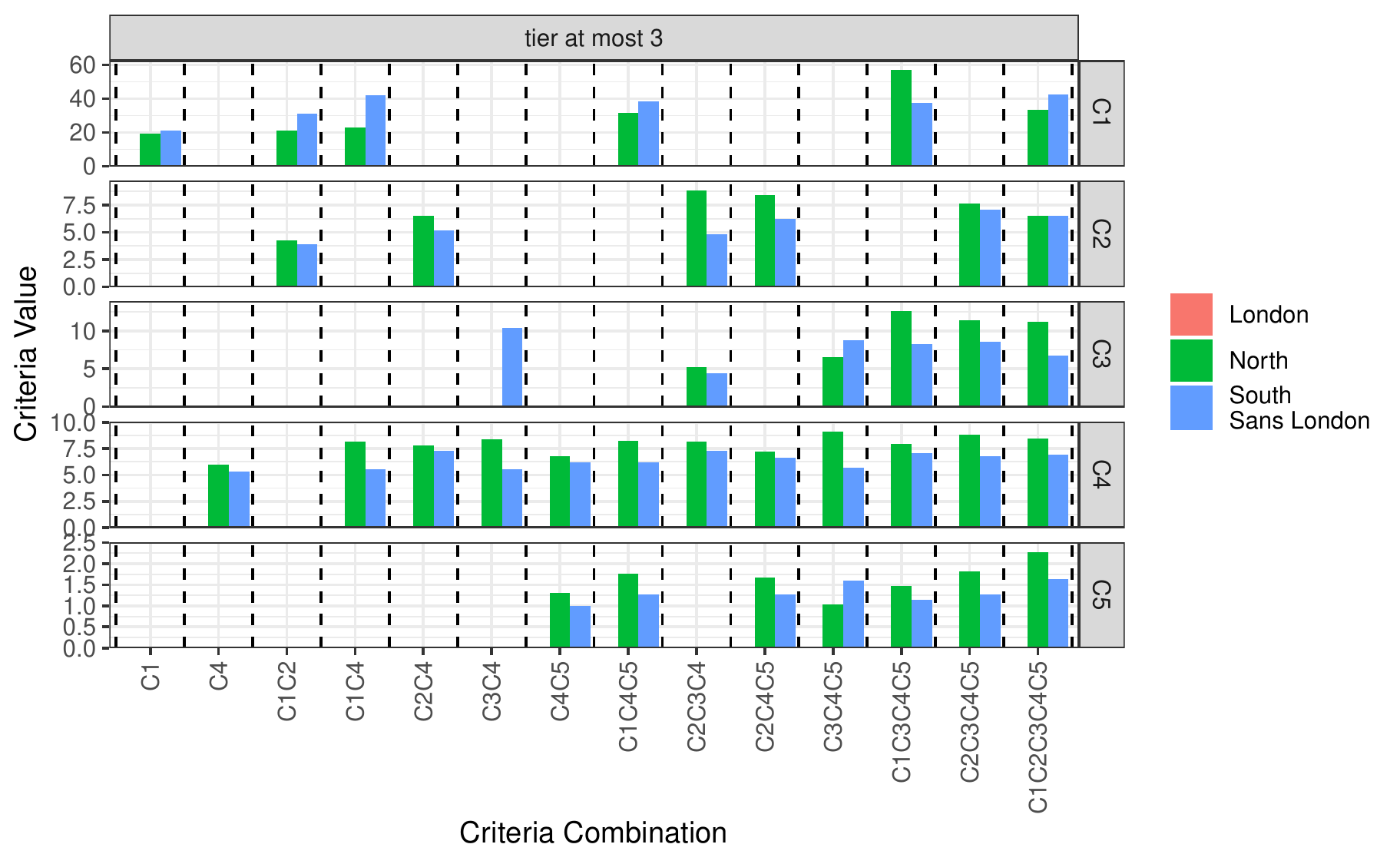}
     \caption{Downward Rules Plot}
     \label{fig:DownwardRules}
\end{figure}

Instead of listing all the rules in a traditional if-then format, we display these rules in a table in order to better compare the three regions (with respect to different outcomes). The extracted upward rules are summarised in Table \ref{tab:UpwardRules}, while the downward rules are summarised in Table \ref{fig:DownwardRules}. One may argue that the upward rules are about restricting regions from entering a lower Tier and so could be considered more relevant from health policy perspective, where as downward rules restrict the regions from entering a higher tier and so are more concerned with the economic considerations.

Taking the first criterion C1 (number of new cases) with consequent "tier at least 2" in the upward rules table (Table \ref{tab:UpwardRules}), the extracted rules suggest that London is at least in Tier 2 for values greater than 297.14, whereas the North is at least in Tier 2 for values greater than 62.04. This implies that the thresholds for London are relatively more relaxed than the North. For example, if the number of new cases is over 62.04 and below 297.14, the extracted rule suggests that North should not be in Tier-1, whilst London might still be in Tier-1. The ratio between these values is 1:4, suggesting London's cases can be almost five fold more before it is treated the same as the North. Note that here there was no such rule extracted for the South to compare with the North and London. 

\begin{theorem}The thresholds for upward rules give evidence of relaxed rules for London for C1\end{theorem}

Focusing on the second criterion of C2 (number of cases in those aged 60+), DRSA has extracted rules for London and the South, however, no rule was extracted for the North. As observed for C1, the threshold for London (i.e. 33.71) are also more relaxed than the threshold for the South (i.e. 14.19). The threshold for London is more than twice the threshold for the South. In this case, there was no rule extracted for the North.

For the fourth criterion of C4 (Positivity rate), DRSA has extracted rules for all three of the geographical regions, facilitating a comparison between all three. Here again, the threshold for London (23.5) is many folds higher than others, whilst there is a negligible difference between the North (6.36) and the South (5.12). 

\begin{theorem}The thresholds for upward rules give evidence of relaxed rules for London for C4 as London was in Tier-2 when others were in Tier-3\end{theorem}

Considering the fifth criterion (Pressure on NHS), again the threshold for London remains very high (4.22), followed by the North (2.09) and then the South (1.00). To summarise, the extracted rules suggest that the thresholds for London are considerably higher than other parts of England.

\begin{theorem}C5 in London remained significantly higher than other regions. \end{theorem}

Focusing on the consequent of tier at least 4, the rules with a single criterion in its antecedent were found only for C4 (see the second last column of Table \ref{tab:UpwardRules}). Here, the rules for the North and the South are showing almost identical threshold values (i.e. 22.67 and 22.63).

So far, we have discussed the rules involving only a single criterion in its antecedent. There also exist other rules involving multiple criteria in their antecedents. For example, the consequent of tier at least 3 shows that there are rules extracted with the criteria C4 and C5 together. The threshold values for these two criteria are 9.09 \& 2.91 for the North, and 13.64 \& 3.3 for the South, respectively. Please note that as these thresholds are paired together, the values cannot be compared individually. In this case, we can see that the pair of values for the North are both considerably smaller than the pair of values extracted for the South.

\begin{table}
     \centering
     \includegraphics[width=1\textwidth]{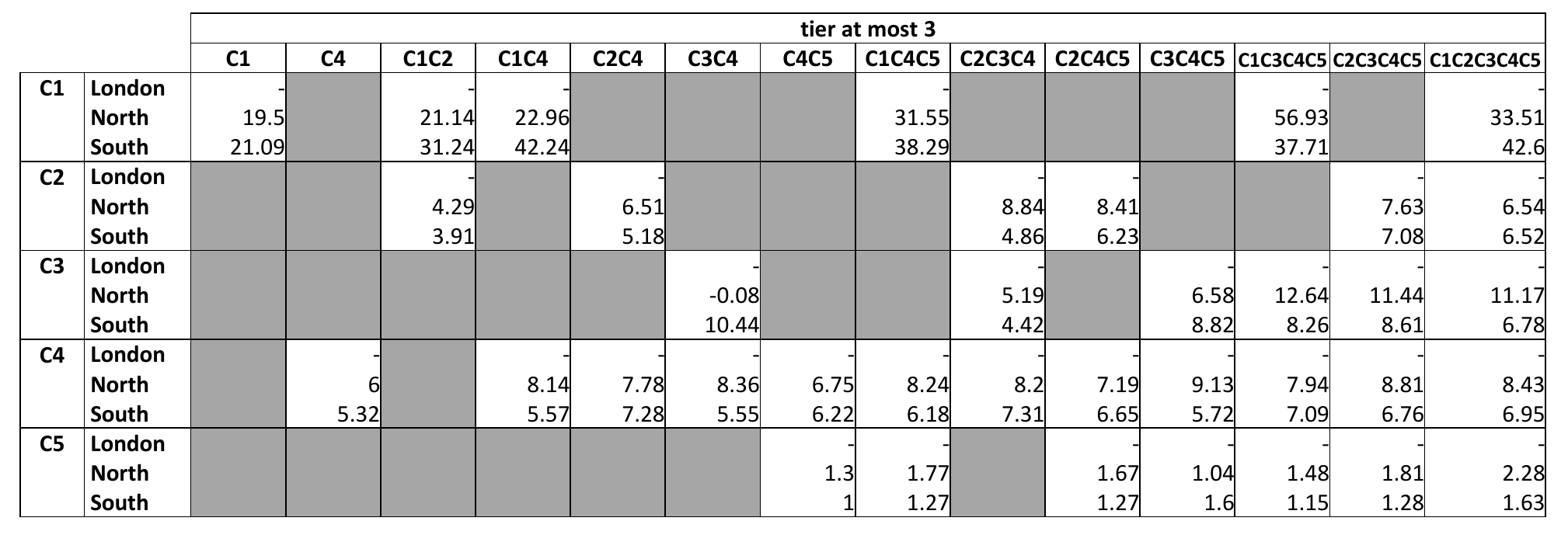}
     \caption{Downward Rules Table}
     \label{tab:DownwardRules}
\end{table}

The downward rules are summarised in Table \ref{tab:DownwardRules}, and are visualised in Figure \ref{fig:DownwardRules} for inspection. Looking at the first row showing thresholds for C1, there are five instances where South sans London has higher threshold values than the North, while there is one instance where North has higher threshold value. On the contrary, looking at the second row, showing thresholds for C2, the North has higher threshold values in all cases. There are no rules extracted for London, however there are several rules extracted for North and South sans London, where, generally there is little to separate the threshold values for these two geographic regions.

Overall, given that the two geographical areas of the North and the South sans London are quite similar, and that London has more relaxed threshold values, it can be argued that although there is a North-South divide in terms of tiers allocation, it is one that is driven by London and not the other regions of the South. This results in London appearing to be given preferential treatment, with respect to the set of Criteria under consideration at least.

With London representing such a large percentage of the economic activity in England, this suggests that there may be additional implicit input into tier allocation reflecting economic considerations as additional criteria, 

\begin{theorem} There is a north south divide, in terms of tiers allocation, which is driven mostly by London.\end{theorem}

\section{Conclusion\label{sec:conclusion}}
In this paper, we investigated the use of fairness in allocating tiered levels of restrictions, with the help of dominance-based rough sets analysis. We identify patterns in the data pertaining to the UK government criteria set for these tiered allocations which were translated into "if-then" type rules. We demonstrate a novel way of investigating fairness by extracting separate sets of rules for separate segments, and then comparing these rule sets to investigate fairness discrepancies in these segments. We found differences in the rules extracted from the North and the South of England, driven in large part by the region of London.

We found that there is a high level of correlation between C1 and C2, from which one could question how useful was it to have both in deciding the tiered restrictions. Intuitively, it makes sense to consider separate criteria to capture age disparities, given the consensus that age has a direct impact upon the severity of illness. However, our analysis suggests that the use of C2 as a separate criterion was statistically redundant and therefore not very useful. Our analysis also suggests that C3 was also not very useful, in a sense that it had almost no explanatory power in allocating tiers.

We argue that, despite being lauded as transparent, the proposed systematic approach was still not transparent enough. The UK government did collect the data on the set of criteria to assign tiers, but the use of this (and any other) information was obfuscated. As discussed earlier, the UK government gave consideration to the local context and exceptional circumstances such as a local but contained outbreak. This information was however not quantifiable and therefore cannot be included in a numerical model, however, it might be a possible area of future work involving the use of text mining techniques.

There is no information made public regarding the relative importance assigned to each of these criteria. As the problem involved multiple criteria, the focus of released information remained on public health which ignored the use of other important information related to, for example, economy, society and technology. Essentially, the allocation of tiers is a problem involving conflicting objectives where minimising the risk conflicts with the maximising of economic prosperity. Therefore, this is an area of research involving trade-off analysis and other multi-criteria decision making techniques.

One of the implications of a perceived lack of transparency of such a system, combined with an apparent disparity in treatment for different segments, could lead to a breakdown of trust. This in turn could lead to non-conformity to the rules, thus defeating the purpose of the whole tiered-allocation system.

\bibliographystyle{elsarticle-harv}

\end{document}